\documentclass[12pt]{article}
\pdfoutput=1
\usepackage{putex}
\usepackage{graphicx}
\usepackage{caption}
\usepackage{subcaption}
\usepackage{epstopdf}
\usepackage{enumerate}
\usepackage{cite}
\usepackage{tensor}
\usepackage{slashed}
\usepackage[aligntableaux=center]{ytableau}
\usepackage[utf8]{inputenc}

\newcommand{\abs}[1]{\left\lvert #1 \right\rvert}

\newcommand {\be} {\begin {equation}}
\newcommand {\ee} {\end {equation}}

\newcommand {\bes} {\begin {equation*}}
\newcommand {\ees} {\end {equation*}}

\newcommand{\es}[2] {\begin{equation} \label{#1} \begin{split} #2 \end{split} \end{equation}}

\newcommand{\Z}{\mathbb{Z}}

\newcommand{\cO}{{\cal O}}

\newcommand{\beq}{\begin{equation}}
\newcommand{\eeq}{\end{equation}}

\def\<{\langle}
\def\>{\rangle}

\begin{document}

\preprint{PUPT-2494}

\institution{PU}{Joseph Henry Laboratories, Princeton University, Princeton, NJ 08544, USA}

\title{Towards Bootstrapping QED$_3$}

\authors{Shai M.~Chester and Silviu S.~Pufu}

\abstract {We initiate the conformal bootstrap study of Quantum Electrodynamics in $2+1$ space-time dimensions (QED$_{3}$) with $N$ flavors of charged fermions by focusing on the 4-point function of four monopole operators with the lowest unit of topological charge. We obtain upper bounds on the scaling dimension of the doubly-charged monopole operator, with and without assuming other gaps in the operator spectrum.   Intriguingly, we find a (gap-dependent) kink in these bounds that comes reasonably close to the large $N$ extrapolation of the scaling dimensions of the singly-charged and doubly-charged monopole operators down to $N=4$ and $N=6$. 
}

\date{}

\maketitle

\tableofcontents
\setlength{\unitlength}{1mm}

\newpage
\section{Introduction and summary}
\label{INTRO}

Quantum electrodynamics in $2+1$ spacetime dimensions (QED$_3$) could be regarded as a toy model for the real world quantum chromodynamics in $3+1$ dimensions because it is an asymptotically free theory that may also exhibit analogs of chiral symmetry breaking \cite{Pisarski:1984dj} and confinement \cite{Polyakov:1975rs,Polyakov:1976fu}.  In a Lagrangian description, the field content of QED$_3$ consists of a $U(1)$ gauge field possibly coupled to several flavors of charged fermions.  When there is no charged matter, the theory confines \cite{Polyakov:1975rs,Polyakov:1976fu}.  When the number $N$ of (two-component complex) fermion flavors is large, it can be argued using $1/N$ perturbation theory that the infrared physics is described by an interacting conformal field theory \cite{Appelquist:1988sr,Nash:1989xx}.   When $N$ is small but non-zero, the precise dynamics remains uncertain, however, because the theory is strongly coupled, and there are only very few non-perturbative tools available.\footnote{Recently, the $\epsilon$-expansion was used to argue that spontaneous chiral symmetry breaking should occur when $N \leq 4$ \cite{DiPietro:2015taa}. The $F$-theorem suggests that it occurs when $N \leq 8$ \cite{Giombi:2015haa}.  Lattice studies suggest that it occurs when $N=2$ \cite{Hands:2004bh,Strouthos:2008kc} or $N=0$\cite{Karthik:2015sgq}, but the situation at larger $N$ is unclear.}  It is believed that in this regime the theory may exhibit analogs of both chiral symmetry breaking and confinement.

In this work, we aim to initiate a study of QED$_3$ at small $N$ using the conformal bootstrap technique \cite{Rattazzi:2008pe}, with the goal of eventually shedding light on the behavior of the theory in this regime.  The conformal bootstrap is a non-perturbative technique that has yielded quite impressive results in other non-supersymmetric examples, such as the 3d Ising model \cite{Kos:2014bka,Kos:2013tga}, the critical $O(N)$ vector model \cite{Kos:2015mba,Kos:2013tga,Chester:2014gqa}, or, more recently, the Gross-Neveu models \cite{Iliesiu:2015qra}, so it is natural to ask whether it can also be used to learn about 3d gauge theories as well.  In its numerical implementation in terms of semi-definite programming, the conformal bootstrap makes use of unitarity and associativity of the operator algebra as applied to 4-point functions of certain operators in a conformal field theory.

In this paper, we assume that the conformal fixed point of QED$_3$ seen in $1/N$ perturbation theory extends to all values of $N$, and study this CFT using the conformal bootstrap.  Explicitly, we derive and study numerically the crossing relations of four monopole operators (to be defined more precisely shortly) for $N=2$, $4$, and $6$.  What we find are rigorous bounds on the scaling dimensions of these monopole operators and of some of the operators appearing in their OPE\@.  We find that these bounds come close to the large $N$ results when extrapolated to small $N$.   In addition, we find certain features in our bounds that are similar to  those that appeared in the bounds of the lowest-dimension operators in 3d CFTs with global $\Z_2$ symmetry when looking at the single 4-point function of $\Z_2$ odd operators.  In that case, examining the crossing equation of a system of mixed correlators yielded an allowed region in the form of an island centered around the 3d Ising CFT\@.  It would be interesting to see if a study of mixed correlators of monopole operators also yields an island-shaped allowed region, though such an analysis is of a numerical complexity beyond what is currently feasible.

Before we delve into the details of our analysis, let us comment on our choice of studying the crossing equations of monopole operators as opposed to those of other operators in the theory.  QED$_3$ with $N$ unit charged fermions $\psi^i$ has $SU(N) \times U(1)$ flavor symmetry.  The fermions transform as a fundamental of $SU(N)$ and are uncharged under $U(1)$.  The monopole operators have non-zero $U(1)$ charge and also transform in fairly complicated representations of $SU(N)$.  In implementing the conformal bootstrap program, one option would have been to consider the 4-point function of the simplest non-monopole scalar operators, the bilinears $\bar \psi_i \psi^j$ transforming in the adjoint of $SU(N)$.  The crossing equations for such a four-point function were worked out in \cite{Berkooz:2014yda}, and it should be straightforward to study the constraints they imply numerically using computer programs such as {\tt SDPB} \cite{Simmons-Duffin:2015qma}.  The disadvantage of studying this four-point function by itself, however, is that besides QED$_3$, there are other theories such as scalar QED, QCD$_3$ or supersymmetric analogs that all have $SU(N)$ flavor adjoint operators with similar properties, and thus from an abstract CFT point of view, it may be hard a priori to distinguish these theories from one another.

What is specific to QED$_3$ and is not shared by its QCD or supersymmetric analogs is indeed the spectrum of monopole operators, and this is why we focus on them.   It can be shown  \cite{Borokhov:2002ib, Dyer:2013fja} that the monopole operator $M_q$ that carries $U(1)$ charge $q \in \Z/2$ also transforms under $SU(N)$ as an irreducible representation given by the Young diagram
 \es{Young}{
    {\tiny N/2} \Bigl\{   \underbrace{ {\tiny \ydiagram{3, 3}} }_{2 \abs{q}} \,.
  }
This feature makes QED$_3$ different from the other similar theories for which the lowest-dimension non-monopole scalars are also $SU(N)$ adjoints.  Note that without any Chern-Simons interactions, $N$ is required to be even in order to avoid a parity anomaly \cite{Borokhov:2002ib}, so the Young diagram \eqref{Young} is indeed well-defined.

Monopole operators are interesting to study not just so that we can distinguish QED$_3$ from other theories.  More generally, they are quite important for the dynamics of gauge theories in $2+1$ dimensions.   The simplest example is pure $U(1)$ gauge theory, where it was shown by Polyakov that their proliferation provides a mechanism for confinement \cite{Polyakov:1975rs}.  If one adds a sufficiently large number $N$ of charged matter fields (bosons or fermions), the infrared physics is believed to be governed by an interacting conformal field theory (CFT), where, in certain condensed matter realizations, monopole operators can act as order parameters for quantum phase transitions that evade the Ginzburg-Landau paradigm \cite{Wen:1993zza, Chen:1993cd, Sachdev97, Rantner01, Rantner:2002zz, Motrunich:2003fz, SVBSF, SBSVF, Hermele, Hermele05, Ran06, Kaul08, Kaul:2008xw, Sachdev:2010uz}.  In these interacting CFTs, the only available method\footnote{For fermions, preliminary $4-\epsilon$ expansion results are discussed in \cite{Chester:2015wao}.} for studying the properties of the monopole operators is the $1/N$ expansion, which so far has been used to compute their scaling dimensions to next-to-leading order in $1/N$ \cite{Murthy:1989ps, Borokhov:2002ib, Metlitski:2008dw, Pufu:2013vpa,Dyer:2013fja,Pufu:2013eda, Dyer:2015zha}.  Going to higher orders in the $1/N$ expansion appears to be very challenging with current techniques. It is nevertheless desirable to learn about monopole operators away from the large $N$ limit, which serves as further motivation for studying them using the conformal bootstrap.

The rest of this paper is organized as follows.  In Section~\ref{review}, we review some known facts about 3d QED and monopole operators.  Sections~\ref{bootstrap} and \ref{numerics} represent the main part of this paper, in the former we compute the crossing equations for the monopole operators in 3d QED, including explicit crossing relations for the cases $N=2\,,4\,,6$, and in the latter we present the results of our numerical bootstrap. In Section \ref{disc} we conclude and discuss further directions. In the Appendix we include the crossing relations for the cases $N=8\,,10\,,12\,,14$.

\section{3d QED and monopole operators}
\label{review}	

The Lagrangian for 3d QED with $N$ complex two-component fermions is
 \es{lagrangian}{
  {\cal L} = - \bar \psi_i \gamma^\mu (\partial_\mu - i A_\mu) \psi^i - \frac{1}{4e^2} F_{\mu\nu} F^{\mu\nu}\,,
 }
where $\psi_i$ are the fermion fields, $A_\mu$ is a $U(1)$ gauge field with field strength $F_{\mu\nu}$, and $e$ is the gauge coupling.  In the following discussion we restrict to the case where $N$ is even so that we may preserve parity and time reversal symmetry \cite{Borokhov:2002ib}.  At large $N$ one can show that this theory flows to an interacting CFT in the infrared where the Maxwell term in \eqref{lagrangian} is irrelevant \cite{Gracey:1993iu,Gracey:1993sn}.   At small $N$ the theory is strongly coupled and difficult to study, although lattice gauge theory studies \cite{Hands:2004bh,Strouthos:2008kc,2009JPhCS.150e2247S} and other arguments \cite{DiPietro:2015taa,Braun:2014wja} suggest that there is a critical value estimated around $N^\text{crit}=2$ below which the theory no longer flows to an interacting CFT\@.

As mentioned in the introduction, in this paper we will work under the assumption that the IR dynamics is governed by a non-trivial interacting CFT whose properties are the same as those derived from the large $N$ expansion extrapolated to finite $N$.  At the CFT fixed point, one can define gauge-invariant order operators built from the fields in the Lagrangian, as well as disorder operators (monopole operators) defined through boundary conditions on these fields.

\subsection{Lowest dimension monopole operators $M_q$}

A monopole operator $M_q$ with topological charge $q$ at the conformal fixed point of 3d QED with $N$ flavors must transform as a representation of the global symmetry group, which includes the conformal group $SO(3,2)$, the flavor symmetry group $SU(N)$, and the $U(1)$ ``topological'' symmetry generated by the topological current
 \es{TopCurrent}{
  J^\text{top}_\mu = \frac{1}{8 \pi} \epsilon_{\mu\nu\rho} F^{\nu\rho} \,,
 } 
which is conserved due to the Bianchi identity obeyed by $F$.  Under the conformal group, $M_q$ has zero spin and scaling dimensions dependent on $q$ and $N$.  See Table~\ref{qTable} for a list of the scaling dimensions $\Delta_{M_q}$ for $q\leq5/2$, as computed for large $N$ \cite{Dyer:2013fja}.   The operator $M_q$ transforms under $SU(N)$ with Young diagram \eqref{Young}. Parity maps monopoles $M_q$ to antimonopoles~$M_{-q}$. 
\begin{table}[!h]
\begin{center}
\begin{tabular}{c||c}
 $\abs{q}$ & $\Delta_{M_q}$ \\
 \hline \hline
 $0$ & $0$ \\
 \hline
 $1/2$ & $0.265 \,N - 0.0383 + O(1/N)$ \\
  \hline
 $1$ & $0.673  \,N  - 0.194 + O(1/N)$ \\
  \hline
 $3/2$ & $1.186  \,N  - 0.422 +O(1/N) $ \\
  \hline
 $2$ & $1.786  \,N - 0.706 + O(1/N) $ \\
  \hline
 $5/2$ & $2.462  \,N - 1.04 + O(1/N) $
\end{tabular}
\caption{Monopole operator dimension $\Delta_{M_q}$ for monopole charge $q$ in $U(1)$ gauge theory with $N$ flavors. \label{qTable}}
\end{center}
\end{table}

In this bootstrap study we consider the four-point function $\langle M_{1/2} M_{-1/2} M_{1/2} M_{-1/2} \rangle$, so we should also review what is known about the conformal primary operators that appear in the OPEs $M_{\pm 1/2}\times M_{\pm 1/2}$ and $M_{\pm 1/2}\times M_{\mp 1/2}$.  These operators can have topological charge $q = \pm 1$ and $q=0$, respectively.  Since under $SU(N)$, $M_{\pm 1/2}$ transform as \eqref{Young}, the operators in both the $M_{\pm 1/2}\times M_{\pm 1/2}$ and $M_{\pm 1/2}\times M_{\mp 1/2}$ OPEs must transform as 
\es{OPEreps}{
\left(1^{N/2}\right)\otimes\left(1^{N/2}\right)=\bigoplus_{n=0}^{N/2} \left(1^{N-2n},2^{n}\right)\,,
}
where $\left(\lambda_1^{\nu_1},\lambda_2^{\nu_2},\dots\right)$ denotes a Young tableau with $\nu_i$ rows of length $\lambda_i$.  There are thus $1 + N/2$ $SU(N)$ irreps in both the $q=\pm 1$ and $q=0$ sectors.  Because of Bose symmetry, only operators with certain spins can appear in each such irrep, as will be discussed in detail in Section~\ref{bootstrap}. In this bootstrap study, we will be interested primarily in bounding the scaling dimension  of the lowest scalar $q=1$ monopole operator $M_1$, which according to \eqref{OPEreps} transforms under $SU(N)$ as $\left(2^{N/2}\right)$. 

\subsection{Lowest dimension scalar $q=0$ operators in OPE $M_{1/2}\times M_{-1/2}$}

In our bootstrap study, it would be useful  to make use of more information on the operators in the $M_{\pm 1/2}\times M_{\pm 1/2}$ and $M_{\pm 1/2}\times M_{\mp 1/2}$ OPEs, such as their scaling dimensions.  

For simplicity, let us focus on the Lorentz scalars with $q=0$ appearing in the $M_{\pm 1/2}\times M_{\mp 1/2}$ OPE\@.  For a given index $n> 0$, for which the $SU(N)$ irrep is $\left(1^{N-2n},2^{n}\right)$, let us denote the lowest dimension primary by ${\cal O}_{n}$, the next lowest by ${\cal O}_{n}'$, and so on.  As mentioned above, all these operators can be built from gauge invariant combinations of $\psi_i$ and $A_\mu$ because they have zero topological charge.  

As will be explained in more detail in \cite{Chester:2016ref}, the operator ${\cal O}_n$ has the form
\es{kOpMain}{
\cO^{}_n=&\psi^{1}_{(\alpha_{1}}\dots\psi^{n}_{\alpha_{n)}}\bar\psi_{n+1}^{(\alpha_{1}}\dots\bar\psi_{2n}^{\alpha_{n})}\,,
}
where $\alpha_m = 1, 2$ are Lorentz spinor indices.  This operator is parity even (odd) depending on whether $n$ is even (odd).  Its scaling dimension is \cite{Chester:2016ref}
\es{246}{
&\Delta_1=2-\frac{64}{3\pi^2N}+O(1/N^2)\,, \qquad
\Delta_2=4-\frac{64}{\pi^2N}+O(1/N^2) \,, \\
&\Delta_3=6-\frac{128}{\pi^2N}+O(1/N^2) \,, 
 \qquad \Delta_4=8-\frac{640}{3\pi^2N}+O(1/N^2) \,,  \\
 &\text{etc.}
}
Note that in this expansion $N$ is taken to infinity before all other quantities.  In particular, the results corresponding to the $n$ channel may break down when $N$ is comparable to $n$.  The next two operators ${\cal O}_{n}'$ and ${\cal O}_{n}''$ have opposite parity from ${\cal O}_{n}$ and can be constructed from $n+1$ $\psi$'s and $n+1$ $\bar \psi$'s.  Their scaling dimensions can also be calculated in the $1/N$ expansion and take the form $\Delta_n' = 2(n+1) + O(1/N)$ and $\Delta_n'' = 2(n+1) + O(1/N)$.

The previous results are only for $n>0$. For $n=0$, i.e.~the $SU(N)$ singlet case, the lowest dimension parity odd operator is $\cO_0\propto\bar\psi_i\psi^i$, whose scaling dimension is given by \cite{Rantner:2002zz}
 \es{DeltaSinglet}{
   \Delta_{0} = 2 + \frac{128}{3 \pi^2 N} + O(1/N^2)  \,.
 }
For the lowest dimension parity even $SU(N)$ singlet, we must consider the mixing between $(\bar \psi_i \psi^i) (\bar \psi_j \psi^j)$ and $F_{\mu\nu}^2$, which gives \cite{Chester:2016ref} 
 \es{ScalingSinglet}{
 \Delta'_{0} =4 + \frac{64 (2 - \sqrt{7}  ) }{3 \pi^2} \frac 1N + O(1/N^2) \,, \qquad
  \Delta''_{0} =4 + \frac{64 (2 + \sqrt{7}  ) }{3 \pi^2} \frac 1N + O(1/N^2) \,.
 }
(See also \cite{2008PhRvB..78e4432X}.)

\subsection{Conserved-current and stress-tensor two-point functions}

Another set of quantities in 3d QED that have been computed in large $N$ are the ``central charges'' $c_T$, $c_J^f$, and $c_J^t$, which are defined as the coefficients of the two-point functions of the conserved stress tensor $T_{\mu\nu}$, $SU(N)$ flavor current $J^f_\mu{}^i{}_j$, and $U(1)$ topological current $J^t_\mu$, respectively, where $J^f_\mu{}^i{}_j$ and $T_{\mu\nu}$ are canonically normalized and $J^t_\mu$ is normalized so that $\int d^2 x\, J^t_0=2q$.\footnote{We have $J^t_\mu = 2 J^{\text{top}}_\mu$, where $J^\text{top}_\mu$ was defined in \eqref{TopCurrent}.}  The two-point functions take the form:
\es{currentNorm}{
\langle J^t_\mu(x)J^t_\nu(0)\rangle&=c_J^{t}\frac{I_{\mu\nu}(x)}{8\pi^2}
     \frac{1}{|x|^{4}} \,,\\
     \langle J^f_\mu{}^i{}_j(x)J^f_\nu{}^k{}_l(0)\rangle&=c_J^{f}\frac{I_{\mu\nu}(x)}{8\pi^2}
     \frac{1}{|x|^{4}}\left(\delta^{i}_l\delta^k_j-\frac{1}{N}\delta^i_j\delta^k_l\right) \,,\\
     \langle T_{\mu\nu}(x)T_{\rho\sigma}(0)\rangle&=c_T\frac{3}{16\pi^2}\left(\frac{1}{2}\left(I_{\mu\rho}(x)I_{\nu\sigma}(x)+I_{\mu\sigma}(x)I_{\nu\rho}(x)\right)-\frac{1}{3}\eta_{\mu\nu}\eta_{\rho\sigma}\right)\frac{1}{\abs{x}^{6}} \,,
}
where $I_{\mu\nu}(x)= \eta_{\mu\nu}-2\frac{x_\mu x_\nu}{x^2}$.\footnote{These definitions are such that $c_T/N = c_J^t/N =c_J^f=1$ for a theory of $N$ free complex two-component fermions in 3d.  In such a theory, $J^f_{\mu}{}^i{}_j$ would be the generator of the $SU(N)$ rotations under which the fermions would transform as a fundamental, and $J^t_\mu$ would be the $U(1)$ current under which all fermions have charge $+1$. \label{cConventions}}

These central charges have been computed to next to leading order in \cite{Giombi:2016fct} as well as \cite{Huh:2013vga,Huh:2014eea}. In our normalization \eqref{currentNorm} we have 
\es{currents}{
c_J^f&\approx 1+\frac{0.1429}{N}+O(1/N^2)\,,\\
c_J^t&\approx\frac{6.4846}{N}-\frac{0.9267}{N^2}+O(1/N^3)\,,\\
c^{}_T/N&\approx 1+\frac{0.7193}{N}+O(1/N^2)\,.
}

\section{Crossing equations}
\label{bootstrap}

We now show how to set up the conformal bootstrap for the four point function of monopole operators in 3d QED\@.   We will focus on the four-point function of two $q=1/2$ monopole operators and two $q=-1/2$ antimonopole operators, which as mentioned previously transform in the $(1^{N/2})$ representation of $SU(N)$, i.e.~they are completely antisymmetric tensors of $SU(N)$ with $N/2$ indices.  Let $M_{1/2}^I$ denote the monopole operator, where $I=\{i_1,\dots,i_{N/2}\}$ and $i=1,\dots,N$ are $SU(N)$ fundamental indices. It is convenient to recast $U(1)$ as $SO(2)$ by writing $M_{1/2}^I=M_{1/2}^{1I}+iM_{1/2}^{2I}$ and $ M_{-1/2}^I=M_{1/2}^{1I}-iM_{1/2}^{2I}$ and working with $M_{1/2}^{aI}$, where $a=1,2$ is a fundamental $SO(2)$ index. We consider the four-point function:
\es{4pt}{
\langle M_{1/2}^{aI}(x_1)M_{1/2}^{bJ}(x_2)M_{1/2}^{cK}(x_3)M_{1/2}^{dL}(x_4)\rangle \,,
}
which includes all orderings of 2 $M_{1/2}$'s and two $M_{-1/2}$'s at once.

The conformal primaries $\cO_{(R,n)}^{\Delta, \ell}$ appearing in the $M_{1/2}^{aI} \times M_{1/2}^{bJ}$ OPE can be classified according to their transformation properties under $SO(2)\times SU(N)$, which are labeled by the index $(R, n)$.  Here, $R$ labels the $SO(2)$ representation, and it can take the values: $R = S$ for $SO(2)$ singlets; $R = A$ for rank-two anti-symmetric tensors\footnote{The singlet ($S$) and rank-two antisymmetric tensor ($A$) representations of $SO(2)$ are of course isomorphic, but it is convenient to keep track of whether ${\cal O}_{(R, n)}^{\Delta, \ell}$ appears in the symmetric ($S$) or anti-symmetric ($A$) product of two $SO(2)$ fundamentals.  As will be explained, the operators in $S$ and $A$ have spins of opposite parity, with those in $S$ having spins of the same parity as that of the operators in $T$.} of $SO(2)$; and $R = T$ for rank-two traceless symmetric tensors.  (In terms of the topological charge $q$, we have that $R = S, A$ correspond to $q=0$ and $R = T$ corresponds to $q = \pm 1$.) For $SU(N)$, we see from \eqref{OPEreps} that we have representations $\left(1^{N-2n},2^n\right)$ where $n=0,\dots,N/2$.    We will show shortly that for each $(R, n)$ only operators with either even $\ell$ or odd $\ell$ can appear in the $M_{1/2}^{aI} \times M_{1/2}^{bJ}$ OPE\@.

Performing the $s$-channel OPE in \eqref{4pt}, we have
 \es{SChannel}{
  \langle M_{1/2}^{aI}(x_1)M_{1/2}^{bJ}(x_2)M_{1/2}^{cK}(x_3)M_{1/2}^{dL}(x_4)\rangle 
   = \sum_{R \in \{S, A, T\}} \sum_{n=0}^{N/2} f_R^{abcd} {\bf t}_n^{IJKL} s_{R, n} 
   \sum_{{\cal O}_{(R, n)}^{\Delta, \ell}} \lambda_{{\cal O}_{(R, n)}^{\Delta, \ell}}^2 g_{\Delta, \ell}(u, v)
 }
where we combined the contribution from each conformal multiplet into a conformal block, and where $f_R^{abcd}$, ${\bf t}_n^{IJKL}$,  $s_{R, n}$, $\lambda_{{\cal O}_{(R, n)}^{\Delta, \ell}}^2$, and $g_{\Delta, \ell}(u, v)$ are defined as follows.  The $f^{abcd}_{R}$ are $SO(2)$ 4-point tensor structures corresponding to exchanging operators in representation $R$ of $SO(2)$.  They are given by \cite{Rattazzi:2010yc}
 \es{fDefs}{
  f^{abcd}_{S} &\equiv \delta^{ab} \delta^{cd} \,, \\
  f^{abcd}_{A} &\equiv \delta^{ad} \delta^{bc} - \delta^{ac} \delta^{bd} \,, \\
  f^{abcd}_{T} &\equiv \delta^{ad} \delta^{bc} + \delta^{ac} \delta^{bd} -\delta^{ab} \delta^{cd} \,.
 } 
The ${\bf t}_n^{IJKL}$ are 4-point tensor structures corresponding to exchanging operators in $\left(1^{N-2n},2^n\right)$ of $SU(N)$.  The $s_{R, n}$ are very important signs ($s_{R, n} = +1$ or $-1$) that are determined by unitarity, as we will discuss in Section~\ref{SO2SUN}.  The $\lambda_{{\cal O}_{(R, n)}^{\Delta, \ell}}^2$ are the squares of the OPE coefficients that must be positive by unitarity.  (We can normalize the OPE coefficient of the identity operator $\lambda_\text{Id}$=1.)  Lastly, $g_{\Delta, \ell}(u, v)$ are conformal blocks corresponding to the exchange of the operator ${\cal O}_{(R, n)}^{\Delta, \ell}$, normalized, for concreteness, as in \cite{Kos:2013tga}.

Swapping $(1,I,a)\leftrightarrow(3,K,c)$ in the four point function \eqref{4pt} yields crossing equations of the form
\es{crossing}{
  \sum_{\cO \in\,M_{1/2}^{aI} \times M_{1/2}^{bJ}} \lambda^2_{\cO}\, \vec{d}^{\,R,n}_{\Delta,\ell}(\Delta_{M_{1/2}},u,v)=0 \,, 
}
where $\cO $ runs over all conformal primaries in the $M_{1/2}^{aI} \times M_{1/2}^{bJ}$ OPE\@.  The crossing function $ \vec{d}^{\,R,n}_{\Delta,\ell}$ is a $3(N/2+1)$ component vector.  (The number of components is determined according to \cite{Rattazzi:2010yc} by the number of $SO(2)\times SU(N)$ representations $R,n$ that occur in the $M_{1/2}^{aI} \times M_{1/2}^{bJ}$ OPE, where representations with both odd and even spins contribute twice.) The components of the crossing function are explicit functions of the conformally-invariant cross-ratios $u = \frac{x_{12}^2 x_{34}^2}{x_{13}^2 x_{24}^2}$ and $v = \frac{x_{14}^2 x_{23}^2}{x_{13}^2 x_{24}^2}$. The form of $ \vec{d}^{\,R,n}_{\Delta,\ell}$ depends only on the dimension of both the external monopole operator $\Delta_{M_{1/2}}$ and on the dimension $\Delta_{\cO}$, Lorentz spin $\ell$, and $SO(2)\times SU(N)$ representation $(R,n)$ of the operator $\cO$. In the rest of this section we provide an efficient algorithm to compute $ \vec{d}^{\,R,n}_{\Delta,\ell}$ for any $N$, which we demonstrate explicitly for the cases $N=2,4,6$. The cases $N=8,10,12,14$ are given in Appendix \ref{moreCross}.

The Lorentz scalars $M_{1/2}^{aI}$ transforms in the fundamental of $SO(2)$ and in the representation $\left(1^{N/2}\right)$ of $SU(N)$. The crossing equations of an operator such as $M_{1/2}^{aI}$ that transforms under a product group can be expressed, roughly, as a tensor product of the crossing equations under each group factor.  In this case, we rewrite \eqref{crossing} more explicitly as 
 \es{eq:crossingWithON}{
   0 &= \sum_{\cO\in n, \ell^+}  \lambda_\mathcal{O}^2 \vec{d}^{\,S,n}_{ \Delta, \ell}
    + \sum_{\cO\in n, \ell^-}  \lambda_\mathcal{O}^2 \vec{d}^{\,A,n}_{ \Delta, \ell}
    +\sum_{\cO\in n, \ell^+}  \lambda_\mathcal{O}^2 \vec{d}^{\,T,n}_{ \Delta, \ell}\,,
 }
where $\vec{d}^{\,R,n}_{ \Delta, \ell}$ are given by the $O(2)$ fundamental crossing functions \cite{Rattazzi:2010yc}
 \es{VDefs}{
\vec{d}^{\,S,n}_{ \Delta, \ell}= s_{S, n} \left(
\begin{matrix}
0\\
\vec{d}^{\,-,n}_{ \Delta, \ell}\\
\vec{d}^{\,+,n}_{ \Delta, \ell}
\end{matrix} \right) \,,
\qquad 
\vec{d}^{\,A,n}_{ \Delta, \ell}= s_{A, n} \left(
\begin{matrix}
-\vec{d}^{\,-,n}_{ \Delta, \ell}\\
\vec{d}^{\,-,n}_{ \Delta, \ell}\\
-\vec{d}^{\,+,n}_{ \Delta, \ell}
\end{matrix} \right),
\qquad  \vec{d}^{\,T,n}_{ \Delta, \ell}= s_{T, n} \left(
\begin{matrix}
\vec{d}^{\,-,n}_{ \Delta, \ell}\\
0\\
-2\vec{d}^{\,+,n}_{ \Delta, \ell}
\end{matrix} \right) \,, 
 }
with $\vec{d}^{\,\pm,n}_{ \Delta, \ell}$ being the crossing functions under $SU(N)$ that we will describe next.  In \eqref{eq:crossingWithON}, the notation $\ell^+$ ($\ell^-$) means that we sum over the same (opposite) set of spins as the component $SU(N)$ crossing functions. 

\subsection{Known results for $N=2, 4$}

In the cases $N=2,4$, the crossing functions $\vec{d}^{\,\pm, n}_{\Delta, \ell}$ appearing in \eqref{VDefs} are already known. When $N=2$, the representation $(1^{N/2}) = (1)$ of the external operator is the fundamental representation of $SU(2)$.  The corresponding crossing functions are a reduced version of the general fundamental $SU(N)$ crossing functions written in \cite{Rattazzi:2010yc}, and they are given by\footnote{We multiplied $\vec{d}_{\Delta,\ell}^{\mp,1}$ by an overall minus sign in order to agree with the conventions we use in Section~\ref{SU2}.  For now, we can think of this minus sign as a redefinition of the $s_{R, 1}$ coefficients in \eqref{VDefs}.  These coefficients will be determined in Section~\ref{SO2SUN}.} 
\es{SU2cross}{
N=2:\qquad\qquad\vec{d}_{\Delta,\ell}^{\mp,0}=\begin{pmatrix}
F_{\Delta,\ell}^\mp   \\
   F_{\Delta,\ell}^\pm
  \end{pmatrix} \,,
\qquad
\vec{d}_{\Delta,\ell}^{\mp,1}=\begin{pmatrix}
 -F_{\Delta,\ell}^\mp   \\
 3 F_{\Delta,\ell}^\pm 
  \end{pmatrix}\,.
}
Here, the operators in the $n=0$ singlet ($n=1$ adjoint) representations can have odd (even) spins, and the functions $F_{\Delta,\ell}^\pm$ are defined in terms of the conformal blocks $g_{\Delta,\ell}(u,v)$, the conformal cross ratios $u = \frac{x_{12}^2 x_{34}^2}{x_{13}^2 x_{24}^2}$ and $v = \frac{x_{14}^2 x_{23}^2}{x_{13}^2 x_{24}^2}$, and the scaling dimension $\Delta_\text{ext}$ of the external operator: 
\es{Fs}{
F^\pm_{\Delta,\ell}(u,v)=v^{\Delta_{\text{ext}}}g_{\Delta,\ell}(u,v)\pm u^{\Delta_{\text{ext}}}g_{\Delta,\ell}(v,u)\,.
}
Recall that the external operator dimension in our case is $\Delta_\text{ext} = \Delta_{M_{1/2}}$.

For $N=4$, the six dimensional $(1^{2})$ representation of $SU(4)$ is isomorphic to the six dimensional fundamental representation of $SO(6)$, so the crossing functions are given by the $O(6)$ fundamental crossing functions \cite{Rattazzi:2010yc}: 
\es{ONcrossing2}{
N=4:\qquad\vec{d}_{\Delta,\ell}^{\mp,0}=\begin{pmatrix}
 0 \\
F_{\Delta,\ell}^\mp   \\
   F_{\Delta,\ell}^\pm
  \end{pmatrix} \,,
  \qquad
\vec{d}_{\Delta,\ell}^{\mp,1}=\begin{pmatrix}
- F_{\Delta,\ell}^\mp   \\
F_{\Delta,\ell}^\mp   \\
 -  F_{\Delta,\ell}^\pm
  \end{pmatrix} \,,
\qquad
\vec{d}_{\Delta,\ell}^{\mp,2}=\begin{pmatrix}
 F_{\Delta,\ell}^\mp   \\
\frac{2}{3}F_{\Delta,\ell}^\mp   \\
- \frac{4}{3}F_{\Delta,\ell}^\pm 
  \end{pmatrix} \,.  
}
Here, the operators in the singlet $n=0$, antisymmetric $n=1$, and traceless symmetric $n=2$ representations of $O(6)$ can have even, odd, and even spins, respectively. 

 For $N\geq6$ there are no results in the literature for the crossing equations, but they can be efficiently derived using the algorithm described below. As a check on our algorithm, we recover the known results given above for $N=2,4$.

\subsection{General algorithm}

We begin by considering the four point function of operators $\cO_I$ where $I=\{i_1,\dots,i_{N/2}\}$ and $i=1,\dots,N$ are $SU(N)$ fundamental indices:
\es{4ptSUN}{
x^{2\Delta_\text{ext}}_{12}x^{2\Delta_\text{ext}}_{34}\langle\cO^I(x_1)\cO^J(x_2)\cO^K(x_3)\cO^L(x_4)\rangle=\sum_{n=0}^{N/2}  \bold t^{IJKL}_n \sum_{{\cal O}_n}   \lambda^2_{\cO_n}  g_{\Delta,\ell}(u,v)\,,
}
where 
$\bold t^{IJKL}_n$ is the four-point tensor structure that corresponds to the exchange of a conformal multiplet whose primary transforms as $\left(1^{N-2n},2^n\right)$ for $n=0,\dots,N/2$, and we will suppress the sets of $SU(N)$ indices $IJKL$ for now on.  Using explicit expressions for $\bold t_n$, it will be straightforward to implement the crossings $(1,I)\leftrightarrow(3,K)$ and $(1,I)\leftrightarrow(2,J)$. The former crossing will give us the crossing functions, while the latter will give us the allowed spins in each representation.

All the indices on the LHS of \eqref{4ptSUN} are fundamentals of $SU(N)$, which implies that $\bold t_n$ can be written as
\es{bbasis}{
\bold t_n= \bold b_{m'} {\bf U}_{m'n} \,,\qquad \bold b_{m'}=\epsilon^{p_1\dots p_N}\epsilon^{p_{N+1}\dots p_{2N}} \,,
}
where $p\in\{i_1,\dots,i_{N/2},j_1,\dots,j_{N/2},k_1,\dots,k_{N/2},l_1,\dots,l_{N/2}\}$ and $\bold b_{m'}$ form a basis for all tensor structures of this form.

Our first step is to exchange $(I)\leftrightarrow(K)$ or  $(I)\leftrightarrow(J)$ for each $\bold b_{m'}$ and express the result as a linear combination of $\bold b_{m'}$'s:
\es{crossingMatrix}{
(\bold b_{m'})_{(I)\leftrightarrow(K)}=\bold b_{n'} {\bf X}_{n'm'} \,,\qquad(\bold b_{m'})_{(I)\leftrightarrow(J)}= \bold b_{n'} {\bf Y}_{n'm'}\,.
}
Our second step is to compute the matrix $\bold U_{m'n}$ that transforms between the bases $\bold t_n$ and $\bold b_{m'}$. For this purpose we will use the $SU(N)$ rank-2 Casimir, which we define in our case as
\es{cas}{
C_2=\left(\left(T^{(1)}\oplus\dots\oplus T^{(N/2)}\right)^2\right) \,,
}
where $\left(T^{(q)}\right)_{i_q}^{j_q}$ are fundamental $SU(N)$ generators for each index $i_q$, so that $C_2$ acts on $SU(N)$ tensors with $N/2$ fundamental indices $i_q$. $C_2$ acts on the (suppressed) first $N$ fundamental $SU(N)$ indices of $\bold b_{m'}$ as
\es{casBFirst}{
C_2\bold b_{m'}=\bold b_{n'}\bold D_{n'm'}.
}
The eigenvectors $(\bold t_{n})_{m'}$ of $\bold D_{n'm'}$ are eigenvectors of $C_2$ 
\es{casB}{
(C_2\bold t_n)_{n'} =\bold D_{n'm'}(\bold t_n)_{m'}=(c_2)_n(\bold t_n)_{n'} \,.
}
The eigenvalues $(c_2)_n$ of an $SU(N)$ tensor in representation $\left(1^{N-2n},2^n\right)$ for $n=0,\dots,N/2$ can be calculated by standard group theory formulae and are given by
\es{eigs}{
(c_2)_n=n(2N+1-n) \,,
}
so that indexing $\bold t_n$ by order of increasing $(c_2)_n$ is consistent with the original definition of $\bold t_n$ in \eqref{4ptSUN}.  Note that each ${\bf t}_n$ as defined above can be multiplied by any real constant and still obeys \eqref{casB}.  Here, we just make a choice of some ${\bf t}_n$ that obey \eqref{casB}.

The transformation matrix $\bold U_{m'n}$ in \eqref{bbasis} between the bases $\bold t_n$ and $\bold b_{m'}$ is then given by
\es{btok}{
\bold U_{m'n}=\left((\bold t_n)_{m'}\right)
}
where we compute $\left((\bold t_n)_{m'}\right)$ in \eqref{casB}.

Putting everything together, the crossing function $\vec d^{-,n}_{\Delta,\ell}$ for the exchange $(1,I)\leftrightarrow(3,K)$ acting on the four point function \eqref{4ptSUN} is an $(N/2+1)\times(N/2+1)$ matrix given by 
\es{crossSUNfunc}{
\vec d^{-,n}_{\Delta,\ell}
   = {\bf b}_{m'} \left[ \bold X_{m'n'}\bold U_{n'n} u^{\Delta_\text{ext}}g_{\Delta,\ell}(v,u)- {\bf U}_{m' n} v^{\Delta_\text{ext}}g_{\Delta,\ell}(u,v) \right]\,,
}
which we can rewrite in terms of $F^\pm_{\Delta,\ell}(u,v)$ using the definition \eqref{Fs}.  When expressing $\vec d^{-,n}_{\Delta,\ell}$ as a column vector, it is convenient to do so in a basis different from ${\bf b}_{m'}$ that is chosen such that some components involve only $F^+_{\Delta, \ell}(u, v)$ and some only $F^-_{\Delta, \ell}(u, v)$.

The analogous equation for the exchange $(1,I)\leftrightarrow(2,J)$, with $\bold X\leftrightarrow\bold Y$, will yield equations of form $F^\pm_{\Delta,\ell}(u,v)\lambda^2_{\cO_n}=0$ for each representation $n$, which for $F^-$, $F^+$ imposes even, odd spins for that representation. 

To demonstrate this algorithm, we will now perform it explicitly for the cases $N=2,4,6$. The crossing functions for $N=8,10,12,14$ are given in Appendix \ref{moreCross}.
\subsubsection{$N=2$}
\label{SU2}

We choose the $\bold b_{m'}$ basis:
\es{bSU2}{
\bold b_0=\epsilon^{ij}\epsilon^{kl}\,,\qquad\bold b_1=\epsilon^{ik}\epsilon^{jl}
}
The exchanges $(I)\leftrightarrow(K)$ or  $(I)\leftrightarrow(J)$ yield the transformation matrices:
\es{xySU2}{
\bold X=\begin{pmatrix}
1&0   \\
 -1& -1
  \end{pmatrix}\,,\qquad \bold Y=\begin{pmatrix}
-1&-1   \\
 0& 1
  \end{pmatrix}\,.
}
Acting with the Casimir $C_2$ on \eqref{bSU2} gives the matrix
\es{DSU2}{
\bold D=\begin{pmatrix}
0&-1   \\
 0& 2
  \end{pmatrix}
}
whose eigenvectors form the matrix
\es{USU2}{
\bold U=\begin{pmatrix}
1&-1   \\
 0& 2
  \end{pmatrix}\,.
}
Quite nicely, this matrix gives the basis of 4-point structures ${\bf t}_0^{ijkl} = \epsilon^{ij} \epsilon^{kl}$, which is anti-symmetric under $i \leftrightarrow j$ and $k \leftrightarrow l$, as appropriate for the exchange of an $SU(2)$ singlet, and ${\bf t}_1^{ijkl} = -\epsilon^{ij} \epsilon^{kl} + 2 \epsilon^{ik} \epsilon^{jl} = \epsilon^{ik} \epsilon^{jl} + \epsilon^{il} \epsilon^{jk} $, which is symmetric under $i \leftrightarrow j$ and $k \leftrightarrow l$, as appropriate for the exchange of an $SU(2)$ triplet.

Constructing the $(1,I)\leftrightarrow(3,K)$ $SU(2)$ crossing function as in \eqref{crossSUNfunc} yields the expected result given in \eqref{SU2cross} for the coefficients of $\vec d^{-,n}_{\Delta,\ell}$ in the basis ${\bf b}'_0 = -{\bf b}_0 + \frac 12 {\bf b}_1$ and ${\bf b}'_1 = - \frac 12 {\bf b}_1$---after rewriting \eqref{crossSUNfunc} in terms of ${\bf b}'_0$ and ${\bf b}'_1$, one can identify the coefficients of ${\bf b}'_0$ with the first row of \eqref{SU2cross} and the coefficients of ${\bf b}'_1$ with the second row of \eqref{SU2cross}.  It can be checked that the $(1,I)\leftrightarrow(2,J)$ $SU(2)$ crossing equations are consistent with the expected spin parities required by Bose symmetry, namely odd and even for $\bold t_0$ (singlet) and $\bold t_1$ (adjoint), respectively.\footnote{Bose symmetry requires that only even (odd) spin operators appear in the symmetric (anti-symmetric) product of the representations of the external operators.  It is not hard to see that the representations with $N-n$ even (odd) appear in the symmetric product of $(1^{N/2})$ with itself, so they should contain operators with even (odd) spins if no other flavor symmetries are present.  If other flavor symmetries are present (such as $SO(2)$ in our case), then the spin parity of the operators for each $n$ is the same as above in the symmetric product of the representations of the other flavor symmetries, and opposite to above in the anti-symmetric product.\label{footnoteParity}}

\subsubsection{$N=4$}
\label{SU4}

We choose the $\bold b_{m'}$ basis:
\es{bSU4}{
\bold b_0=\epsilon^{i_1i_2j_1j_2}\epsilon^{k_1k_2l_1l_2}\,,\qquad\bold b_1=\epsilon^{i_1i_2k_1k_2}\epsilon^{j_1j_2l_1l_2}\,, \qquad
\bold b_2=\epsilon^{i_1i_2l_1l_2}\epsilon^{j_1j_2k_1k_2}\,.
}
The exchanges $(I)\leftrightarrow(K)$ or  $(I)\leftrightarrow(J)$ yield the transformation matrices:
\es{xySU4}{
\bold X=\begin{pmatrix}
0&0 &1  \\
 0& 1&0\\
 1& 0&0\\
  \end{pmatrix}\,,\qquad \bold Y=\begin{pmatrix}
1&0 &0  \\
 0& 0&1\\
 0& 1&0\\
  \end{pmatrix}\,.
}
Acting with the Casimir $C_2$ on \eqref{bSU2} gives the matrix
\es{DSU4}{
\bold D=\begin{pmatrix}
0&-1 &-1  \\
 0& 5&1\\
 0& 1&5\\
  \end{pmatrix}
}
whose eigenvectors form the matrix 
\es{USU4}{
\bold U=\begin{pmatrix}
1&0 &-\frac 13  \\
 0& -1&1\\
 0& 1&1\\
  \end{pmatrix}\,.
}
Constructing the $(1,I)\leftrightarrow(3,K)$ $SU(4)$ crossing function as in \eqref{crossSUNfunc} yields the expected result for $\vec{d}^{-, n}_{\Delta, \ell}$ given in \eqref{ONcrossing2}, when expressing the components of $\vec{d}^{-, n}_{\Delta, \ell}$ in the basis ${\bf b}'_0 = -{\bf b}_1$, ${\bf b}'_1 = -({\bf b}_0 + {\bf b}_2)/2$, ${\bf b}'_2 = -({\bf b}_0 - {\bf b}_2)/2$.  The $(1,I)\leftrightarrow(2,J)$ $SU(4)$ crossing equations are consistent with the expected spin parities required by Bose symmetry: even, odd, even for $\bold t_0$ (singlet), $\bold t_1$ (adjoint/antisymmetric), $\bold t_2$ (symmetric), respectively---see Footnote~\ref{footnoteParity}.

\subsubsection{$N=6$}
\label{SU6}

We choose the $\bold b_{m'}$ basis:
\es{bSU6}{
&\bold b_0=\epsilon^{i_1i_2i_3j_1j_2j_3}\epsilon^{k_1k_2k_3l_1l_2l_3}\,,\qquad\bold b_1=\epsilon^{i_1i_2i_3k_1k_2k_3}\epsilon^{j_1j_2j_3l_1l_2l_3}\,,\\
&\bold b_2=\epsilon^{i_1i_2i_3l_1l_2l_3}\epsilon^{j_1j_2j_3k_1k_2k_3}\,,\qquad\bold b_3=\epsilon^{i_1i_2i_3j_1j_2k_1}\epsilon^{j_3k_2k_3l_1l_2l_3}\,.
}
The exchanges $(I)\leftrightarrow(K)$ or  $(I)\leftrightarrow(J)$ yield the transformation matrices: 
\es{xySU6}{
\bold X=\begin{pmatrix}
0&0 &-1& -2 \\
 0& -1&0&-1\\
 -1& 0&0&-2\\
 0&0&0&1\\
  \end{pmatrix}\,,\qquad \bold Y=\begin{pmatrix}
-1&0 &0&-3  \\
 0& 0&1&0\\
 0& 1&0&0\\
 0&0&0&1
  \end{pmatrix}\,.
}
Acting with the Casimir $C_2$ on \eqref{bSU6} gives the matrix
\es{DSU6}{
\bold D=\begin{pmatrix}
0&1 &2&-9  \\
 0& 11&1&0\\
 0& 1&11&0\\
 0&-1&-1&6
  \end{pmatrix}
}
whose eigenvalues form the matrix
\es{USU6}{
\bold U=\begin{pmatrix}
1&-3 &1&-3  \\
 0& 0&-10&-6\\
 0& 0&10&-6\\
 0&2&0&2
  \end{pmatrix}\,.
}
Constructing the $(1,I)\leftrightarrow(3,K)$ $SU(6)$ crossing function as in \eqref{crossSUNfunc} yields: 
\es{SU6cross}{
N=6:\qquad& \vec{d}^{\,\mp,0}_{\Delta,\ell}= \begin{pmatrix}
0 \\
F_{\Delta,\ell}^\mp   \\
F_{\Delta,\ell}^\pm   \\
 0
  \end{pmatrix}\,,
\qquad
 \vec{d}^{\,\mp,1}_{\Delta,\ell}=  \begin{pmatrix}
F_{\Delta,\ell}^\mp   \\
0  \\
0  \\
F_{\Delta,\ell}^\pm
  \end{pmatrix}\,,\\
& \vec{d}^{\,\mp,2}_{\Delta,\ell}= \begin{pmatrix}
0  \\
-9F_{\Delta,\ell}^\mp   \\
21F_{\Delta,\ell}^\pm   \\
 - 10 F_{\Delta,\ell}^\pm
  \end{pmatrix}\,,
  \qquad
  \vec{d}^{\,\mp,3}_{\Delta,\ell}= \begin{pmatrix}
 F_{\Delta,\ell}^\mp   \\
6F_{\Delta,\ell}^\mp   \\
0\\
 - 5 F_{\Delta,\ell}^\pm
  \end{pmatrix}\,.
}
Here, the components of the column vectors are the components of $\vec{d}^{\mp, n}_{\Delta, \ell}$ in the basis ${\bf b}'_0 = (7 {\bf b}_0 + 2 {\bf b}_1 + {\bf b}_2 - 4 {\bf b}_3)/2$, ${\bf b}'_1 = -({\bf b}_0 - {\bf b}_2)/2$, ${\bf b}'_2 = -({\bf b}_0 + {\bf b}_2)/2$, ${\bf b}'_3 = -({\bf b}_0 + 2 {\bf b}_1 +  {\bf b}_2)/2$.  The $(1,I)\leftrightarrow(2,J)$ $SU(6)$ crossing equations are consistent with the expected  spin parities required by Bose symmetry: odd, even, odd, even for $\bold t_0$ (singlet), $\bold t_1$ (adjoint), $\bold t_2$, and $\bold t_3$, respectively---see Footnote~\ref{footnoteParity}.

\subsection{Reflection Positivity}
\label{SO2SUN}

Reflection positivity is the Euclidean version of the unitarity constraints on a Lorentzian CFT\@. These constraints fix the sign of $\lambda^2_{\cO}$, by demanding that when we consider the four-point function of scalar operators $\langle\cO_1\cO_2\cO_2^\dagger\cO_1^\dagger\rangle$, the coefficients multiplying the conformal blocks in the $s$-channel OPE should be positive \cite{Rattazzi:2008pe}. $SU(N)$ has complex generators, so to enforce this condition in our case, we must define what we mean by the complex conjugate of an operator $\cO^{aI}$ transforming under $SO(2)\times SU(N)$.  In fact, we will consider $\cO^{aI}$ to be real under this notion of complex conjugation.

The subtlety in defining the reality properties of our operators comes from the fact that the $SU(N)$ irrep $(1^{N/2})$ under which these operators transform is real when $N/2$ is even and pseudo-real when $N/2$ is odd.   We thus have two different reality conditions depending on whether $N/2$ is even or odd:
\es{reality}{
N/2 \quad\text{Odd}:\quad(\cO^{aI})^\dagger=& \frac{1}{(N/2)!}\epsilon_{IJ}\epsilon^{ab}\cO^{bJ} \,, \\
N/2 \quad\text{Even}:\quad(\cO^{aI})^\dagger=& \frac{1}{(N/2)!}\epsilon_{IJ}\delta^{ab}\cO^{bJ}  \,, \\
}
where $\epsilon^{IJ}\equiv \epsilon^{i_1 \dots i_{N/2} j_1 \dots j_{N/2}}$.  The overall coefficient as well as the dependance on whether $N/2$ is even or odd in \eqref{reality} can be determined (up to a sign) from the requirement that $(\cO^{aI})^{\dagger\dagger}=\cO^{aI}$.    These reality conditions together with the reflection positivity requirement $\langle {\cal O}(x) {\cal O}^\dagger(-x) \rangle > 0$ imply that we can normalize our operators ${\cal O}^{aI}$ to have the following 2-point functions:
\es{reality2point}{
N/2 \quad\text{Odd}: \langle {\cal O}^{aI}(x_1) {\cal O}^{bJ}(x_2)  \rangle  =&\epsilon^{IJ}\epsilon^{ab}\frac{1}{|x_{12}|^{2\Delta_\cO}} \,, \\
N/2 \quad\text{Even}: \langle {\cal O}^{aI}(x_1) {\cal O}^{bJ}(x_2)  \rangle=&\epsilon^{IJ}\delta^{ab}\frac{1}{|x_{12}|^{2\Delta_\cO}}\,.  
}

There are several ways of determining the signs $s_{R, n}$ appearing in \eqref{VDefs}.    We choose to do so by looking at an example, namely the one where ${\cal O}^{aI}$ represent free fields obeying \eqref{reality2point} with $\Delta_{\cal O} = 1/2$.  In this free theory, the four-point function can be obtained from Wick contractions using \eqref{reality2point}:
 \es{FourPtFree}{
 & \langle {\cal O}^{aI}(x_1) {\cal O}^{bJ}(x_2) {\cal O}^{cK}(x_3) {\cal O}^{dL}(x_4) \rangle_\text{free}= \\
&N/2 \quad\text{Odd}:\quad \frac{1}{x_{12} x_{34}} \left[{\bf b}_0 \epsilon^{ab} \epsilon^{cd} + \sqrt{u} {\bf b}_1  \epsilon^{ac} \epsilon^{bd} + \frac{\sqrt{u}}{\sqrt{v}} {\bf b}_2  \epsilon^{ad} \epsilon^{bc} \right] \,,\\
&N/2 \quad\text{Even}:\quad  \frac{1}{x_{12} x_{34}} \left[{\bf b}_0 \delta^{ab} \delta^{cd} + \sqrt{u} {\bf b}_1  \delta^{ac} \delta^{bd} + \frac{\sqrt{u}}{\sqrt{v}} {\bf b}_2  \delta^{ad} \delta^{bc} \right] \,,
 }
where $\bold b_0,\bold b_1,\bold b_2$ are defined as 
\es{bdefined}{
&\bold b_0=\epsilon^{i_1\dots i_{N/2}j_1\dots j_{N/2}}\epsilon^{k_1\dots k_{N/2}l_1\dots l_{N/2}}\,,\\
&\bold b_1=\epsilon^{i_1\dots i_{N/2}k_1\dots k_{N/2}}\epsilon^{j_1\dots j_{N/2}l_1\dots l_{N/2}}\,,\\
&\bold b_2=\epsilon^{i_1\dots i_{N/2}l_1\dots l_{N/2}}\epsilon^{j_1\dots j_{N/2}k_1\dots k_{N/2}}\,.\\
}

We should express this four-point function in terms of the $SO(2)$ four-point structures \eqref{fDefs} using
 \es{epsProdTof}{
  \epsilon^{ab} \epsilon^{cd} &= -f_{A}^{abcd}\,,\qquad\qquad\qquad\qquad\delta^{ab} \delta^{cd} = f_{S}^{abcd} \,, \\
  \epsilon^{ac} \epsilon^{bd} &= \frac{f_{S}^{abcd} - f_{A}^{abcd} - f_{T}^{abcd} }{2}\,,\qquad \delta^{ac} \delta^{bd} = \frac{f_{S}^{abcd} - f_{A}^{abcd} + f_{T}^{abcd} }{2} \,, \\
  \epsilon^{ad} \epsilon^{bc} &= \frac{f_{S}^{abcd} + f_{A}^{abcd} - f_{T}^{abcd} }{2}\,,\qquad \delta^{ad} \delta^{bc}= \frac{f_{S}^{abcd} + f_{A}^{abcd} + f_{T}^{abcd} }{2} \,.
 }

Plugging \eqref{epsProdTof} into \eqref{FourPtFree}, we obtain
 \es{FourPtFreeAgain}{
 & \langle {\cal O}^{aI}(x_1) {\cal O}^{bJ}(x_2) {\cal O}^{cK}(x_3) {\cal O}^{dL}(x_4) \rangle_\text{free}= \\
&N/2 \quad\text{Odd}:\quad \frac{1}{x_{12} x_{34}} \left[
    \frac{  \frac{{\bf b}_2}{\sqrt{v}}  + {\bf b}_1}{2} \sqrt{u} \left( f_S - f_T \right) 
    +  \left( \frac{\frac{{\bf b}_2}{\sqrt{v}}  - {\bf b}_1 }{2 }\sqrt{u} - {\bf b}_0 \right)  f_A  \right] \,,\\
&N/2 \quad\text{Even}:\quad  \frac{1}{x_{12} x_{34}} \left[  \left( \frac{\frac{{\bf b}_2}{\sqrt{v}}  + {\bf b}_1 }{2 }\sqrt{u} + {\bf b}_0 \right)  f_S 
 +   \left(\frac{  \frac{{\bf b}_2}{\sqrt{v}}  - {\bf b}_1}{2}\right) \sqrt{u}  f_A + \left(\frac{  \frac{{\bf b}_2 }{\sqrt{v}} + {\bf b}_1}{2}\right) \sqrt{u}  f_T  \right] \,.
 }
Finally,  we change from the $\bold b_{m'}$ to the $\bold t_m$ basis using the inverse of the transformation matrix $\bold U$ \eqref{btok}, and expand \eqref{FourPtFreeAgain} in terms of the lowest conformal blocks, using the relations
\es{rtheta}{
&1/\sqrt{v}=1+4\eta r+8\eta^2 r^2+4(4\eta^3-\eta)r^3+O(r^4)\,,\quad \sqrt{u}=4r-8\eta r^2+4(4\eta^2-1)r^3+O(r^4)\,,\\
& g_{0,0}=1+O(r^2)\,,\quad g_{1,0}=r+O(r^3)\,,\quad g_{2,1}=r^2\eta+O(r^4) \,,\quad g_{3,2}=\frac{r^3}{2}(3\eta^2-1)+O(r^5)\,.
}
where $r,\eta$ are functions of $u,v$ defined in \cite{Hogervorst:2013sma}.  We can now read off the signs multiplying the conformal blocks of each tensor structure from this example.  These signs must be the same in all theories where the reality conditions \eqref{reality} are satisfied. We now carry out this program explicitly for the cases $N=2\,,4\,,6$.

\subsubsection{$SU(2)$}

 Computing the inverse of $\bold U$ for $SU(2)$ \eqref{USU2} we get
\es{SU2bTok}{
  \bold b_0 &= \bold t_0 \,, \\
  \bold b_1 &= \frac{\bold t_0 + \bold t_1}{2} \,. \\
   \bold b_2 &=\bold b_1-\bold b_0= \frac{-\bold t_0 + \bold t_1}{2} \,, \\
  }
  where the third equation follows as an identity.
So
 \es{SU2b3Plusb2}{
\frac{\bold b_2}{  \sqrt{v} } \pm \bold b_1 = -\frac{ \bold t_0}{2} \left[ \frac{1}{\sqrt{v}} \mp 1 \right] 
   + \frac{\bold t_1 }{2}\left[ \frac{1}{\sqrt{v}} \pm 1 \right]  \,.
 }

Using the relations \eqref{rtheta}, we express the four point function \eqref{FourPtFreeAgain} for the $N/2$ odd case in terms of conformal blocks:
 \es{SU2FourPtFreeConf}{
  \text{4-pt} &= \frac{1}{x_{12} x_{34}} \biggl[
     \left(  2g_{1, 0} \bold t_1- 4 g_{2, 1} \bold t_0 + \cdots \right)  \left( f_S - f_T \right)  \\
    &+  \left(-\bold t_0 g_{0, 0} -  2g_{1, 0}  \bold t_0 + 4 g_{2, 1} \bold t_1  -4g_{3,2}\bold t_0+ \cdots \right)  f_A  \biggr] \,.
 }
Table~\ref{SU2spinTable} follows.  As a consistency check, the spin parities in this table match our computation in Section~\ref{SU2}.

\begin{table}[htp]
\begin{center}
\begin{tabular}{|c|c|c|c|}
 \hline
 $SU(2)$ & $SO(2)$ & spin & $s_{R, n}$ \\
 \hline\hline
 $0$ & $S$ & odd & $-1$ \\
 $1$ & $S$ & even & $1$ \\
 \hline
 $0$ & $A$ & even & $-1$ \\
 $1$ & $A$ & odd & $1$ \\
 \hline
 $0$ & $T$ & odd & $1$ \\
 $1$ & $T$ & even & $-1$ \\
 \hline
\end{tabular}
\end{center}
\caption{Properties of conformal blocks and signs $s_{R, n}$ from \eqref{VDefs} for the case $N=2$.}
\label{SU2spinTable}
\end{table}%

\subsubsection{$SU(4)$}

 Computing the inverse of $\bold U$ for $SU(4)$ \eqref{USU4} we get 
\es{SU4bTok}{
  \bold b_0 &= \bold t_0 \,, \\
  \bold b_1 &= \frac{\bold t_0}{6}  -\frac{\bold t_1 - {\bf t}_2}{2} \,, \\
  \bold b_2 &= \frac{\bold t_0}{6} +\frac{\bold t_1 + {\bf t}_2}{2}  \,.
 }
So
 \es{SU4b3Plusb2}{
  \frac{\bold b_2}{\sqrt{v}}  \pm \bold b_1 = \frac{\bold t_1}{2}\left[ \frac{1}{\sqrt{v}} \mp 1 \right] 
   + \frac{3\bold t_2 + \bold t_0}{6}\left[ \frac{1}{\sqrt{v}} \pm 1 \right]  \,.
 }

Using the relations \eqref{rtheta}, we express the four point function \eqref{FourPtFreeAgain} for the $N/2$ odd case in terms of conformal blocks:
 \es{SU4FourPtFreeConf}{
  \text{4-pt} &= \frac{1}{x_{12} x_{34}} \biggl[
     \left( \bold t_0 g_{0,0} +2g_{1,0}\frac{\bold t_0+3 \bold t_2}{3}+4\bold t_1 g_{2,1} + 4g_{3,2}\frac{\bold t_0+3 \bold t_2}{3}+\cdots \right)   f_S    \\
    &+  \left(2\bold t_1 g_{1, 0} +  4 g_{2,1}  \frac{3 \bold t_2 + \bold t_0}{3}+ \cdots \right)  f_A + \left( 2g_{1,0}\frac{\bold t_0+3 \bold t_2}{3}+4\bold t_1 g_{2,1} + \cdots \right)   f_T  \biggr] \,.
 }
Table~\ref{SU4spinTable} follows.  As a consistency check, the spin parities in this table match our computation in Section~\ref{SU4}.

\begin{table}[htp]
\begin{center}
\begin{tabular}{|c|c|c|c|}
 \hline
 $SU(4)$ & $SO(2)$ & spin & $s_{R, n}$ \\
 \hline\hline
 $0$ & $S$ & even & $1$ \\
 $1$ & $S$ & odd & $1$ \\
 $2$ & $S$ & even & $1$ \\
 \hline
 $0$ & $A$ & odd & $1$ \\
 $1$ & $A$ & even & $1$ \\
 $2$ & $A$ & odd & $1$ \\
 \hline
 $0$ & $T$ & even & $1$ \\
 $1$ & $T$ & odd & $1$ \\
 $2$ & $T$ & even & $1$ \\
 \hline
\end{tabular}
\end{center}
\caption{Properties of conformal blocks and signs $s_{R, n}$ from \eqref{VDefs} for the case $N=4$.}
\label{SU4spinTable}
\end{table}%

\subsubsection{$SU(6)$}

 Computing the inverse of $\bold U$ for $SU(6)$ \eqref{USU6} we get
\es{bTok}{
  \bold b_0 &= \bold t_0 \,, \\
  \bold b_1 &= -\frac{\bold t_2 - \bold t_0}{20}  + \frac{\bold t_1 - \bold t_3}{12} \,, \\
  \bold b_2 &= \frac{\bold t_2 - \bold t_0}{20} + \frac{\bold t_1 - \bold t_3}{12} \,.
 }
So
 \es{b3Plusb2}{
\frac{\bold b_2}{  \sqrt{v}}  \pm \bold b_1 = \frac{\bold t_2 - \bold t_0}{20} \left[ \frac{1}{\sqrt{v}} \mp 1 \right] 
   + \frac{\bold t_1 - \bold t_3}{12}\left[ \frac{1}{\sqrt{v}} \pm 1 \right]  \,.
 }

Using the relations \eqref{rtheta}, we express the four point function \eqref{FourPtFreeAgain} for the $N/2$ odd case in terms of conformal blocks:
 \es{FourPtFreeConf}{
  \text{4-pt} &= \frac{1}{x_{12} x_{34}} \biggl[
     \left(  g_{1, 0} \frac{\bold t_1 - \bold t_3}{3} + 2 g_{2, 1} \frac{\bold t_2 - \bold t_0}{5} + \cdots \right)  \left( f_S - f_T \right)  \\
    &+  \left(-\bold t_0 g_{0, 0} +   g_{1, 0}  \frac{\bold t_2 - \bold t_0}{5}+ 2 g_{2, 1}  \frac{\bold t_1 - \bold t_3}{3} +2 g_{3, 2}  \frac{\bold t_2 - \bold t_0}{5} + \cdots \right)  f_A  \biggr] \,.
 }
Table~\ref{spinTable} follows.  As a consistency check, the spin parities in this table match our computation in Section~\ref{SU6}.

\begin{table}[htp]
\begin{center}
\begin{tabular}{|c|c|c|c|}
 \hline
 $SU(6)$ & $SO(2)$ & spin & $s_{R, n}$ \\
 \hline\hline
 $0$ & $S$ & odd & $-1$ \\
 $1$ & $S$ & even & $1$ \\
 $2$ & $S$ & odd & $1$ \\
 $3$ & $S$ & even & $-1$ \\
 \hline
 $0$ & $A$ & even & $-1$ \\
 $1$ & $A$ & odd & $1$ \\
 $2$ & $A$ & even & $1$ \\
 $3$ & $A$ & odd & $-1$ \\
 \hline
 $0$ & $T$ & odd & $1$ \\
 $1$ & $T$ & even & $-1$ \\
 $2$ & $T$ & odd & $-1$ \\
 $3$ & $T$ & even & $1$ \\ 
 \hline
\end{tabular}
\end{center}
\caption{Properties of conformal blocks and signs $s_{R, n}$ from \eqref{VDefs} for the case $N=6$.}
\label{spinTable}
\end{table}%

\subsection{Constraints From Space-Time Parity}
\label{parity}

As described in \cite{Borokhov:2002ib}, space-time parity maps a monopole operator $M_q$ to an anti-monopole operator with opposite charge $ M_{-q}$. In terms of $SO(2)$ indices, parity acts by sending $1\to1$ and $2\to-2$, thus the $S$ sector is parity even, the $A$ sector is parity odd, and the $T$ sector can transform as both even or odd for different operators.

To find the parity of the uncharged spin 0 operators in each $SU(N)$ sector, we must determine whether they are in the $A$ or $S$ sector. Operators appearing in the $M_{1/2}^{aI} \times M_{1/2}^{bJ}$ OPE\@ have even/odd spins depending on whether they appear in the symmetric/antisymmetric product of the combined $SO(2)\times SU(N)$ representation. Thus operators in representations $S,T$ and $n = N/2, N/2 - 2, \ldots$ or $A$ and $n = N/2-1, N/2 - 3, \ldots$ all have even spin, while the rest have odd spin. The parity of spin 0 uncharged operators must therefore be even for $n = N/2, N/2 - 2, \ldots$, and odd for $n = N/2-1, N/2 - 3, \ldots$\,.

 As described in Section~\ref{review}, the two lowest dimension spin 0 operators $\cO_n$ and $\cO'_n$ in $SU(N)$ representations $\left(1^{N-2n},2^n\right)$ are composed of $2n$ and $2n+2$ fermions, respectively. The parity of a $2n$ fermion operator is even/odd for $n$ even/odd, so the lowest dimension spin 0 operator $\cO_{(n)}$ in $SU(N)$ representations $\left(1^{N-2n},2^n\right)$ with the required parity depends on whether $N/2$ is even or odd. In Table~\ref{parityTable} we show which operator $\cO_n$ or $\cO'_n$ is the lowest dimension operator with the required parity for each $SU(N)$ sectors for $N/2$ even or odd. The scaling dimensions of these operators presented in Section~\ref{review} will be used to motivate the gaps we impose in the subsequent Section~\ref{bounds}.
 
 \begin{table}[htp]
\begin{center}
  \begin{tabular}{c|c|c}
       $n=$ & $N/2-0,2,\dots$      & $N/2-1,3,\dots$  \\
        \hline\hline
$N/2$ Even & $\cO_n$&   $\cO_n'$ \\
$N/2$ Odd  &  $\cO_n'$ &   $\cO_n$  
  \end{tabular}
  \end{center}
  \caption{Composite fermion operator in representation $\left(1^{N-2n},2^n\right)$ with required parity for $N/2$ even or odd.\label{parityTable}}
\end{table}

\section{Numerical bounds}
\label{numerics}
\subsection{Strategy}

After deriving the precise form of the crossing equations \eqref{crossing}, in order to find bounds on the scaling dimensions of operators appearing in the $M_{1/2}^{aI} \times M_{1/2}^{bJ}$ OPE, one can consider linear functionals $\alpha$ satisfying the following conditions:
\begin{align}
1.&\qquad\alpha(\vec{d}_{\text{Id}})=1 \,,\label{1}\\
2.&\qquad\alpha\left( \vec{d}^{\,R,n}_{\Delta,\ell}\right)\geq0, \quad \text{for all $\Delta \geq\Delta_{R,n, \ell}^*$} \label{2}
\end{align}
where $\Delta^*_{R,n, \ell}$ are the assumed lower bounds for spin-$\ell$ conformal primaries (other than the identity) that appear in the $M_{1/2}^{aI} \times M_{1/2}^{bJ}$ OPE and transform in the $SO(2)\times SU(N)$ representation $(R, n)$. The existence of any such $\alpha$ would contradict \eqref{crossing}, and thereby would allow us to find an upper bound on the lowest-dimension $\Delta^*_{R,n, \ell}$ of the spin-$\ell$ conformal primary in representation $R,n$. In particular, if we set $\Delta^*_{T,N/2, 0} = \Delta_{M_{1}}$ and all other $\Delta^*_{R,n, \ell}$ equal to either their unitarity value or some gap value, then we can then find a disallowed region in the $(\Delta_{M_{1/2}},\Delta_{M_1})$ plane for our chosen gap assumptions. 

The above procedure allows us to put gaps for operators that do not have both the same representation and spin as the operator we are bounding. If we would like to put a gap above the operator $\cO_{(R',n'),\ell'}$ that we are bounding, then we must add the following condition:
\begin{align}
3.&\qquad\alpha\left(\vec{d}^{\,R',n'}_{\Delta',\ell'}(\Delta_{M_{1/2}})\right)\geq0 \,,
\end{align}
as well as make sure in condition \eqref{2} that $\Delta^*_{R',n',\ell'}>\Delta'_{R',n',\ell'}$. 

To find lower bounds on the central charges of conserved currents, we relate these charges to OPE coefficients of conformal primaries appearing in the $M_{1/2}^{aI} \times M_{1/2}^{bJ}$ OPE, for which we can find upper bounds using the bootstrap.  On general grounds, the relation must take the form
\es{ctoAnsatz}{
c_J^{t} \propto \frac{\lambda^2_{R, 0, 0, 0}}{\lambda^2_{R,0,2,1}}\,,\qquad c_J^{f}\propto \frac{\lambda^2_{R, 0, 0, 0}}{\lambda^2_{R,1,2,1}}\,, \qquad
c_T\propto \frac{\Delta_{M_{1/2}}^2 \lambda^2_{R, 0, 0, 0}}{\lambda^2_{R,0,3,2}}\,,
}
where the OPE coefficient $\lambda_{R,n,\Delta,\ell}$  has $R$ either $S$ or $A$ depending on which $SO(2)$ representation gives the prescribed spin for the given $SU(N)$ representation, and $n=0\,,1$ are the singlet, adjoint representations of $SU(N)$.  The OPE coefficient $\lambda^2_{R, 0, 0, 0}$ of the identity operator can be chosen to be equal to $1$ as a normalization condition for the external operator.  The coefficients of proportionality in \eqref{ctoAnsatz} can be found from the free theory presented in Section~\ref{SO2SUN}.  A theory of free scalars transforming in representation $R$ of $SU(N)$ and fundamental representation of $SO(2)$, with the reality condition \eqref{reality} has\footnote{The definitions of the central charges are those of footnote~\ref{cConventions} or Eq.~\eqref{currentNorm}.}
 \es{centralFree}{
  c_T^\text{free} = c_J^{t, \text{free}} = \dim R\,,\qquad
   c_J^{f, \text{free}} = \frac{2  C_2(R) \dim R}{N^2 - 1} \,,
 }
where $\dim R$ is the dimension of $R$ and $C_2(R)$ is the value of the quadratic Casimir of the representation.  For us, $R = (1^{N/2})$, which has $C_2(R) = N(N+1)/8$ and $\dim R = \binom{N}{N/2}$.  Comparing these values with the explicit four-point function decompositions in \eqref{SU2FourPtFreeConf}, \eqref{SU4FourPtFreeConf}, and \eqref{FourPtFreeConf}, we find
\es{ctoOPE}{
c_J^{t}=\frac{8\lambda^2_{R, 0, 0, 0}}{\lambda^2_{R,0,2,1}}\,,\qquad c_J^{f}=\frac{A_N \lambda^2_{R, 0, 0, 0}}{\lambda^2_{R,1,2,1}}\,, \qquad
c_T=\frac{32\Delta_{M_{1/2}}^2\lambda^2_{R, 0, 0, 0}}{\lambda^2_{R,0,3,2}}\,,
}
with $A_2 = 4$, $A_4 = 8$, and $A_6 = 2$.  

Using \eqref{ctoOPE}, the lower bounds on the central charges can be recast as upper bounds on certain OPE coefficients.  Upper bounds on the OPE coefficient of an operator $\cO^*$ can be determined by considering linear functionals $\vec{\alpha}$ satisfying the following conditions:
\begin{align}
1.&\qquad\alpha(\vec{d}_{\cO^*})=1 \,,\label{21}\\
2.&\qquad\alpha\left( \vec{d}^{\,R,n}_{\Delta,\ell}\right)\geq0, \quad \text{for all $\Delta \geq\Delta_{R,n, \ell}^*$} \label{22}
\end{align}
where $\Delta^*_{R,n, \ell}$ are the assumed lower bounds for spin-$\ell$ conformal primaries (other than the identity) that appear in the $M_{1/2}^{aI} \times M_{1/2}^{bJ}$ OPE and transform in the $SO(2)\times SU(N)$ representation $R$. If such a functional $\vec{\alpha}$ exists, then this $\alpha$ applied to \eqref{crossing} along with the positivity of all $\lambda_{\cO}^2$ except, possibly, for that of $\lambda_{\cO^*}^2$ implies that
 \es{UpperOPE}{
  \lambda_{\cO^*}^2 \leq - \lambda_\text{Id}^2 \alpha ( \vec{d}_\text{Id}) 
   }
provided that the scaling dimensions of each $\cO\neq\cO$ satisfies $\Delta \geq \Delta_{R,n,\ell}^*$. We can choose the spectrum to only satisfy unitarity bounds, or impose gaps on various sectors. To obtain the most stringent upper bound on $\lambda_{\cO^*}^2$, and therefore lower bound on its associated central charges, one should then minimize the RHS of \eqref{UpperOPE} under the constraints \eqref{22}.

The numerical implementation of the above problems requires two truncations: one in the number of derivatives used to construct $\alpha$ and one in the range of spins $\ell$ that we consider, whose contributions to the conformal blocks are exponentially suppressed for large $\ell$.  We denote the maximum derivative order by $\Lambda$ (as in \cite{Chester:2014fya}) and the maximum spin by $\ell_\text{max}$.  The truncated constraint problem can be rephrased as a semidefinite programing problem using the method developed in \cite{Rattazzi:2008pe}. This problem can than be solved efficiently using {\tt sdpb} \cite{Simmons-Duffin:2015qma}. In this study, we set $\Lambda=19$ and $\ell_\text{max}=25$. We checked that increasing $\Lambda$ and $\ell_\text{max}$ did not change the values of $\Delta_{M_{1/2}}$ or $\Delta_{M_{1}}$ by more than $.01$ for $N=2,4$, and $.02$ for $N=6$. In terms of computing time, {\tt sdpb} took approximately 4 cpu hour for $N=2$, 12 cpu hours for $N=3$, and 18 cpu hours for $N=6$.

\subsection{Numerical bounds for $N=2,4,6$}
\label{bounds}

We now present bounds on scaling dimensions and central charges using the numerical conformal bootstrap. The number of crossing equations, and therefore the numerical complexity, increases as $3(N/2+1)$, so we will only focus on the cases $N=2,4,6$. We use the crossing functions and spin parities computed in the previous section. We will also impose gaps on operators in the uncharged $U(1)$ sector, motivated by the operator scaling dimensions in Section~\ref{review}. The parity constraints discussed in Section~\ref{parity} require that for $N=2,6$ the lowest dimension operators in $SU(N)$ representation $\left(1^{N-n},2^{2n}\right)$ are the $(2n+2)$-fermion operators of dimension $\Delta'_n$, while for $N=4$ they are the $2n$-fermion operators of dimension $\Delta_n$. In the singlet $n=0$ sector, $N=2,6$ has the $2$-fermion operator of dimension $\Delta_0$, while $N=4$ has the $4$-fermion operator of dimension $\Delta'_0$. As the $1/N$ expansion for these values still seems rather large for $N=2,3,4$, the precise numerical values obtained from the large $N$ expansion will serve more as rough guides than exact inputs. 

\subsubsection{Bounds on $\Delta_{M_1}$}
\label{M1bounds}

\begin{figure}[t!]
\begin{center}
   \includegraphics[width=0.49\textwidth]{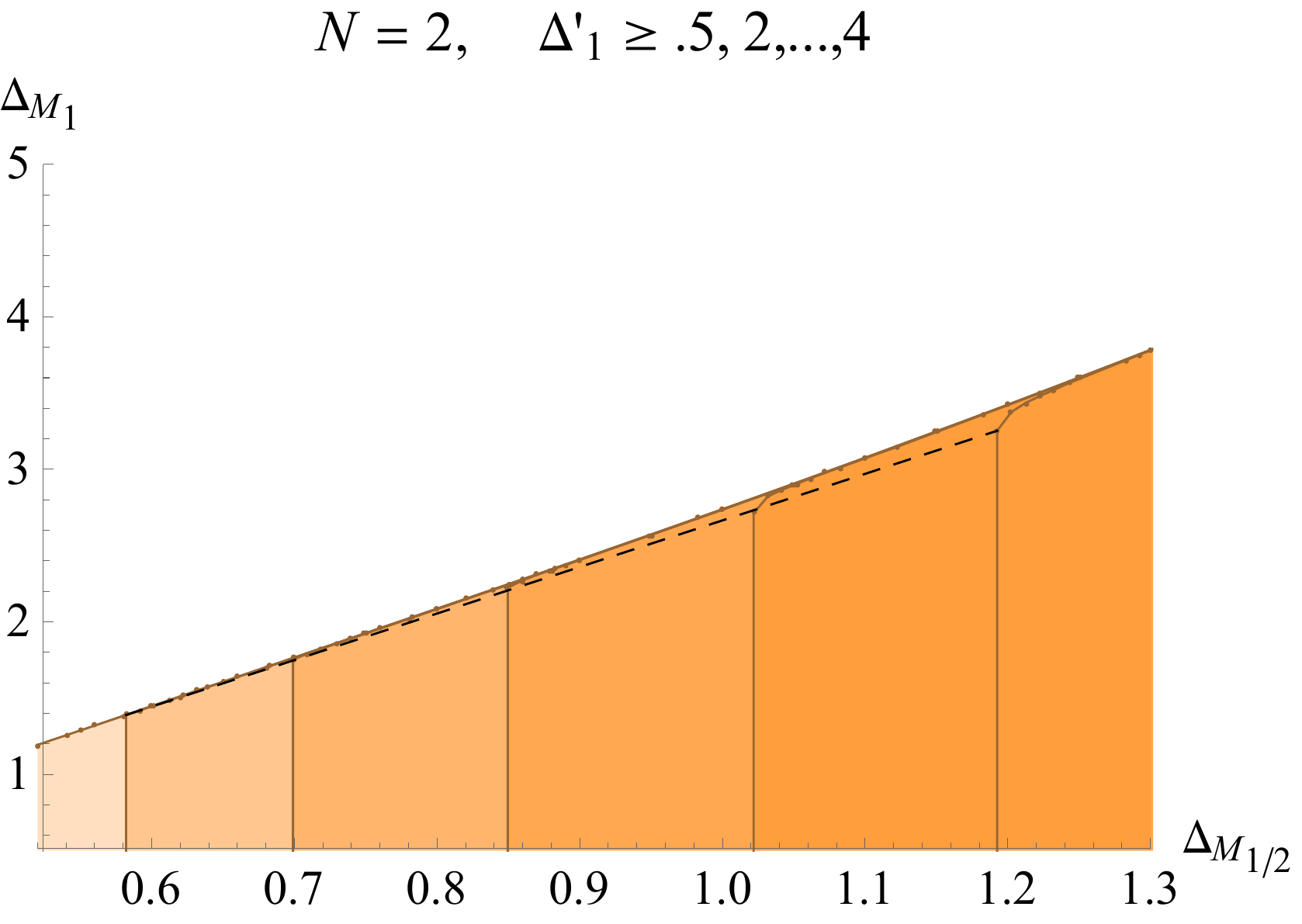}
    \includegraphics[width=0.5\textwidth]{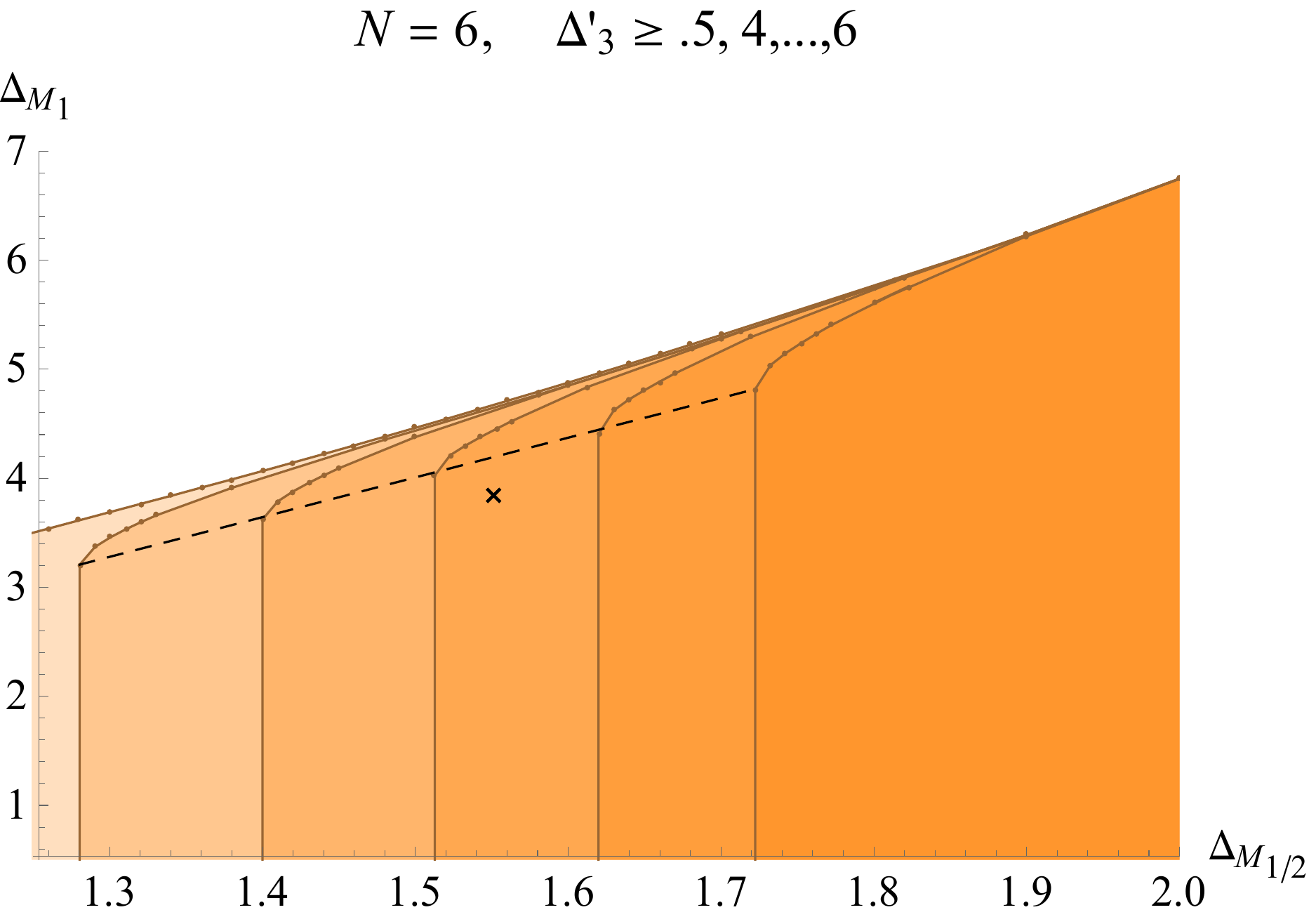}
\caption{
Bounds on basic $q=1$ monopole operator scaling dimension $\Delta_{M_1}$ in terms of basic $q=1/2$ monopole operator scaling dimension $\Delta_{M_{1/2}}$ in $d=3$ for $N=2,6$ (left,right) with gaps $\Delta'_1\geq.5,2,2.5,3,3.5,4$ for $N=2$ and $\Delta'_3\geq.5,2,2.5,3,3.5,4$ for $N=6$ in the uncharged sector in the same $SU(N)$ representation $\left(2^{N/2}\right)$ as $M_1$. These bounds were computed with $\ell_\text{max}=25$ and $\Lambda=19$. The black cross denotes the large $N$ expansion values of $(\Delta_{M_{1/2}},\Delta_{M_1})$.}
\label{fig:26}
\end{center}
\end{figure}

\begin{figure}[t!]
\begin{center}
   \includegraphics[width=0.49\textwidth]{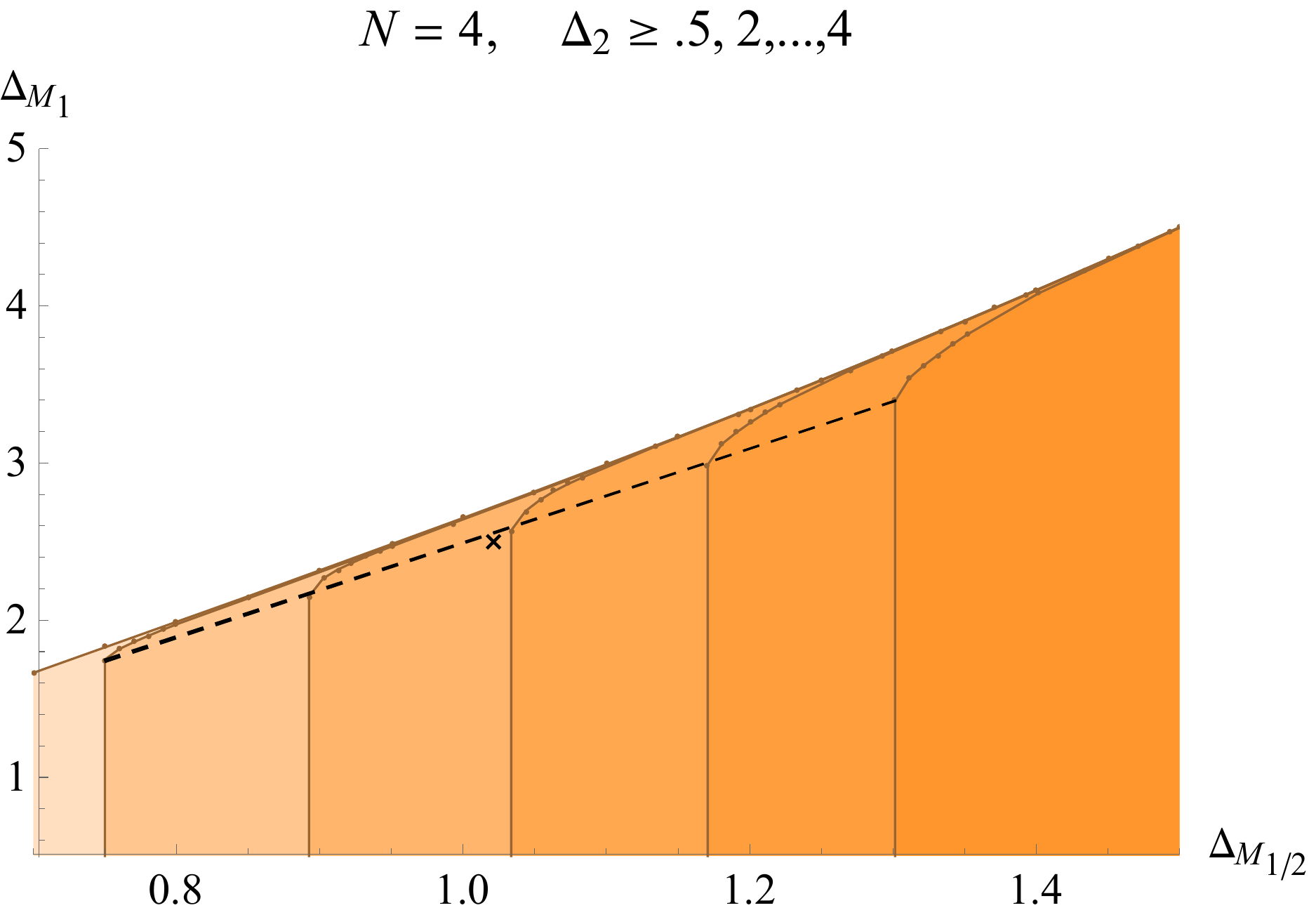}
    \includegraphics[width=0.5\textwidth]{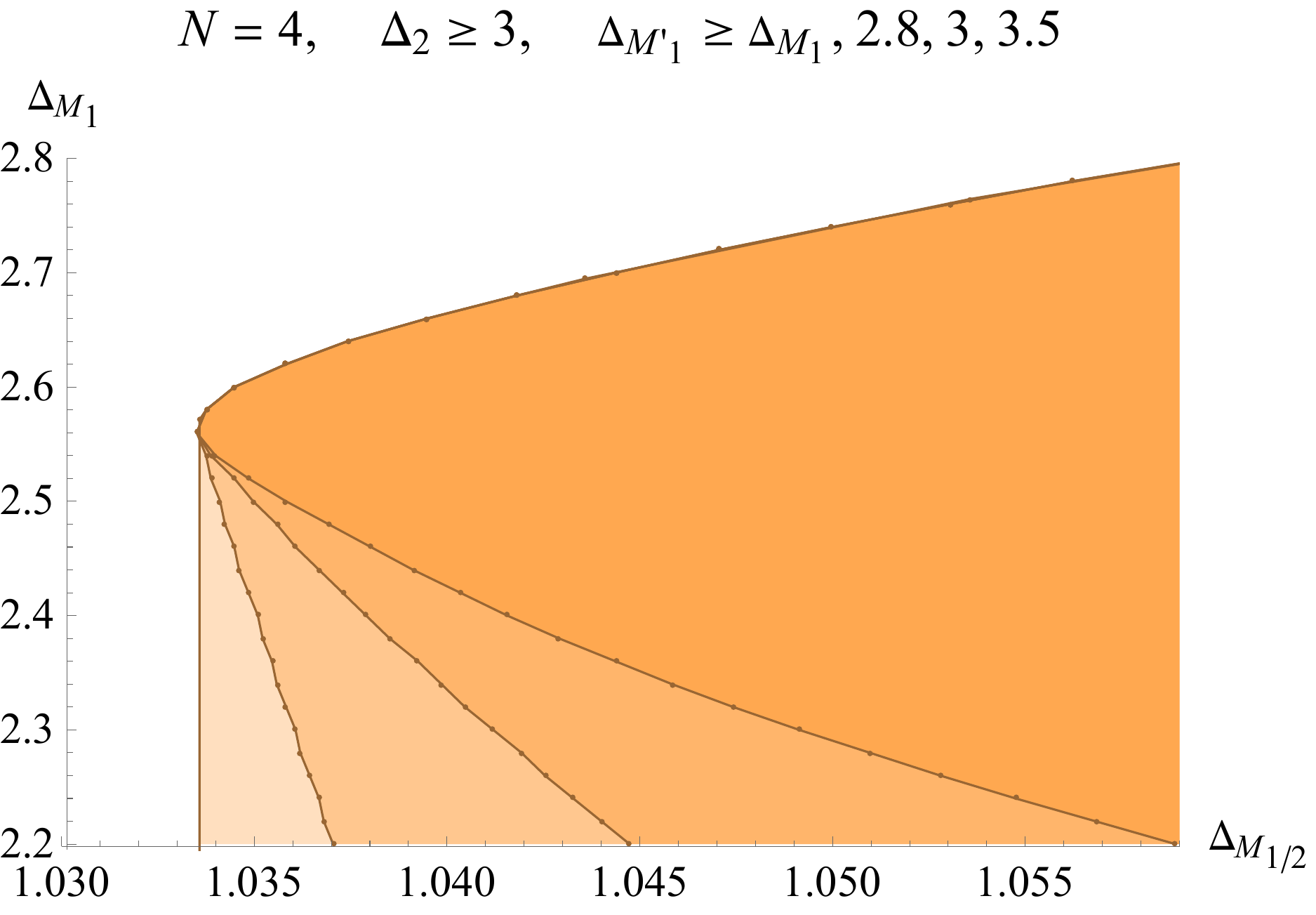}
\caption{Bounds on basic $q=1$ monopole operator scaling dimension $\Delta_{M_1}$ in terms of basic $q=1/2$ monopole operator scaling dimension $\Delta_{M_{1/2}}$ in $d=3$ for $N=4$ with gaps $\Delta_2\geq.5,2,2.5,3,3.5,4$ in the uncharged sector in the same $SU(N)$ representation $\left(2^{N/2}\right)$ as $M_1$. The righthand plot focuses on the $\Delta_2\geq3$ case, and shows that placing an additional gap $\Delta_{M'_1}$ above $\Delta_{M_1}$ creates a peninsula around the kinks seen in the lefthand plots. These bounds were computed with $\ell_\text{max}=25$ and $\Lambda=19$. The black cross denotes the large $N$ expansion values of $(\Delta_{M_{1/2}},\Delta_{M_1})$.}
\label{fig:4}
\end{center}
\end{figure}

In Figures~\ref{fig:26} and~\ref{fig:4} we show bounds on the basic $q=1$ monopole operator scaling dimension $\Delta_{M_1}$ in terms of basic $q=1/2$ monopole operator scaling dimension $\Delta_{M_{1/2}}$.   As seen from Figure~\ref{fig:26} and the left plot of Figure~\ref{fig:4}, when a sufficiently large gap is imposed in the uncharged $U(1)$ sector in the same $SU(N)$ representation $\left(2^{N/2}\right)$ as $M_1$, then a lower bound on $\Delta_{M_{1/2}}$ with an associated $\Delta_{M_1}$ value appears. This feature (kink) seems to depend linearly on this gap---see the dotted lines in Figures~\ref{fig:26} and~\ref{fig:4}. Moreover, the slope of this line of kinks has the same value, $\approx3$, for all of the values of $N$ that we considered. It is a reassuring check on our crossing equations, which differ drastically in form, that all these plots show the same qualitative features.

For the cases $N=4,6$ we mark the large $N$ prediction listed in Table \ref{qTable} for $(\Delta_{M_{1/2}},\Delta_{M_1})$ with a cross in the corresponding plots.\footnote{For $N=2$ this value lies below unitarity, which could mean that there is no corresponding CFT\@.} For $N=4$, the large $N$ extrapolation seems to lie almost exactly on the dotted line connecting the kinks, which implies that a certain value of the gap $\Delta_{2}$ will give a feature at exactly the predicted value in the $(\Delta_{M_{1/2}},\Delta_{M_1})$ plane. We note that imposing reasonable gaps\footnote{For instance, $\Delta'_0\approx3.5$ in the $n=0$ sector and $\Delta_1\approx1.5$ in the $n=1$ sector, as suggested by the large $N$ expressions \eqref{246} and \eqref{ScalingSinglet}, respectively.} in the other uncharged sectors for $N=4$ does not noticeably change the plots. For $N=6$, the large $N$ value lies somewhat below the dotted line connecting the kinks.   We found that for $N=6$, unlike $N=4$, imposing gaps in the other uncharged sectors does change the location of the kinks and brings the line joining the kinks down closer to the large $N$ extrapolation value.

In Figure \ref{fig:4}, the righthand plot focuses on the gap $\Delta_2=3$ case, which from the lefthand plot seems to match the large $N$ values of $(\Delta_{M_{1/2}},\Delta_{M_1})$ best. The righthand plot puts an additional gap $\Delta_{M'_{1}}$ above $\Delta_{M_{1}}$.  We find that any value of $\Delta_{M'_{1}}>\Delta_{M_{1}}$ creates a peninsular allowed region around the kink seen in the lefthand plot. In previous bootstrap studies \cite{Kos:2015mba, Kos:2014bka}, it was found that such a peninsula leads to islands once mixed correctors are used---see, for instance, Figure~3 in \cite{Kos:2014bka}.  It would be interesting to see whether a similar phenomenon occurs here.

\subsubsection{Bounds on $c_T$, $c_J^f$, $c_J^t$}
\label{cbounds}

\begin{figure}[ht!]
\begin{center}
 \includegraphics[width = 0.49\textwidth]{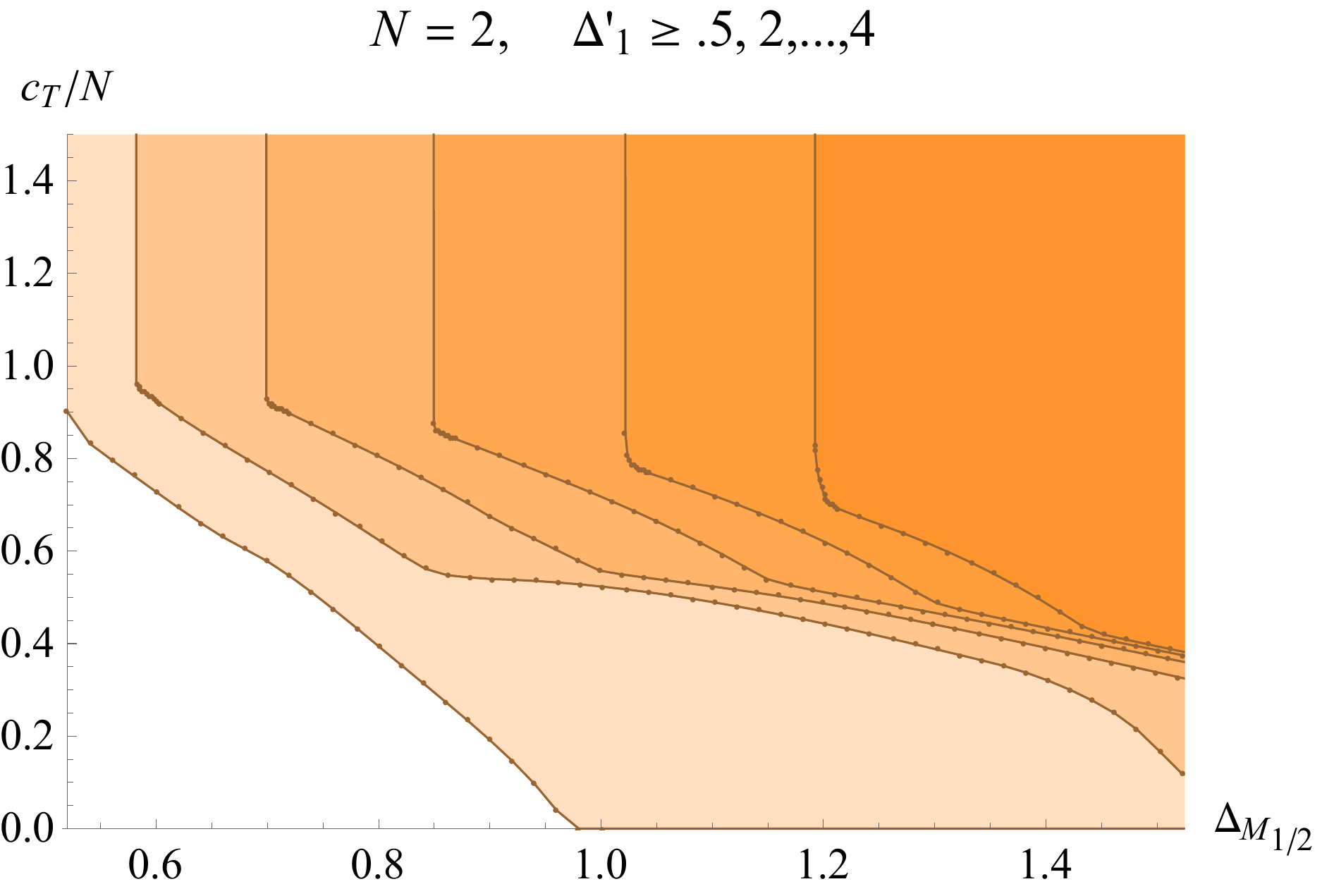}
 \includegraphics[width = 0.49\textwidth]{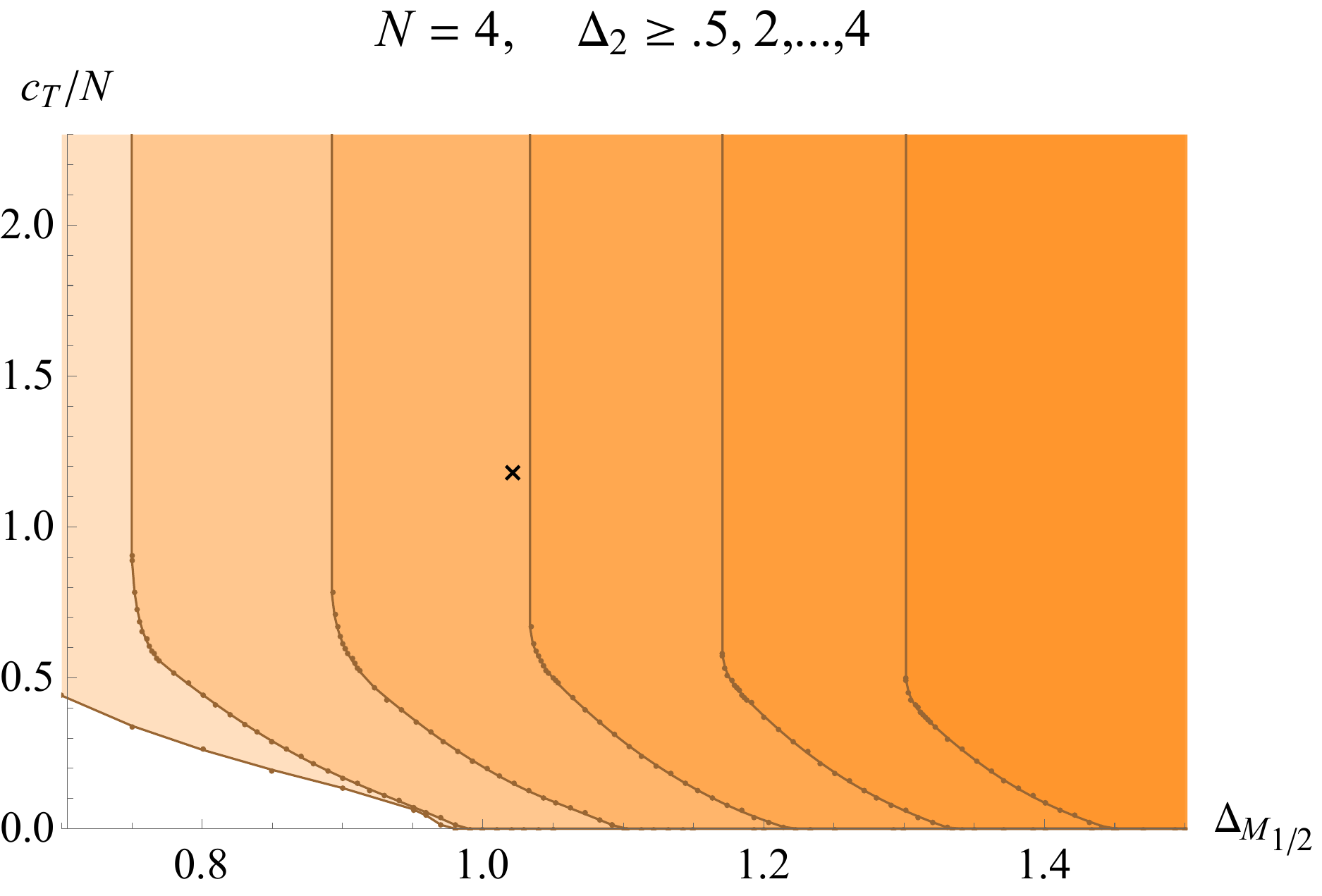} \\
 \includegraphics[width = 0.49\textwidth]{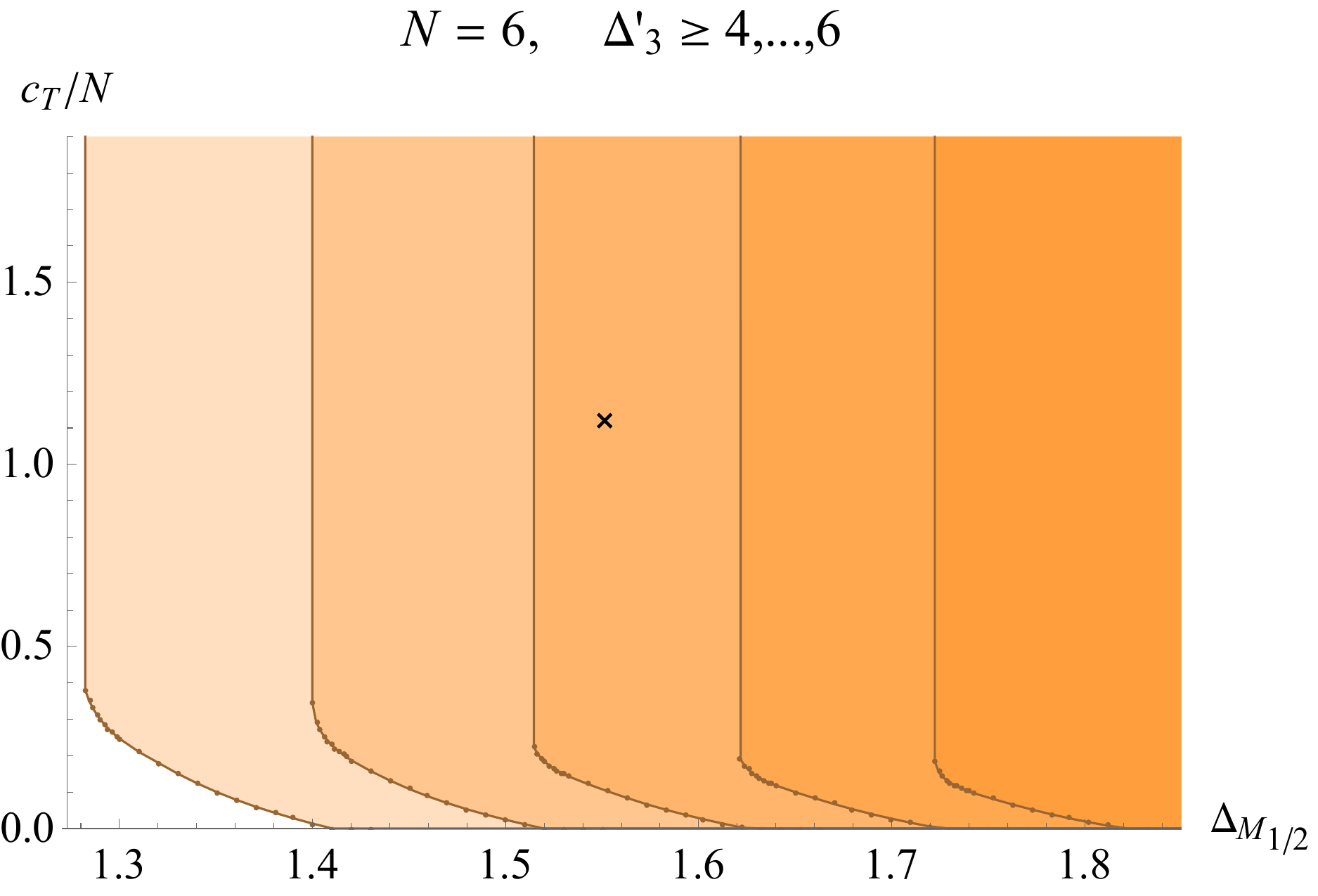}
 \caption{Bounds on stress tensor central charge $c_T$ in terms of basic $q=1/2$ monopole operator scaling dimension $\Delta_{M_{1/2}}$ in $d=3$ for $N=2,4,6$ with gaps $\Delta'_2\geq.5,2,2.5,3,3.5,4$ for $N=2$, $\Delta_4\geq.5,2,2.5,3,3.5,4$ for $N=4$, and $\Delta'_6\geq2,2.5,3,3.5,4$ for $N=6$ in the uncharged sector in the same $SU(N)$ representation $\left(2^{N/2}\right)$ as $M_1$. These bounds were computed with $\ell_\text{max}=25$ and $\Lambda=19$. The black crosses denote the large $N$ expansion values of $c_T$.  \label{fig:cT}}
 \end{center}
 \end{figure}

\begin{figure}[ht!]
\begin{center}
 \includegraphics[width = 0.49\textwidth]{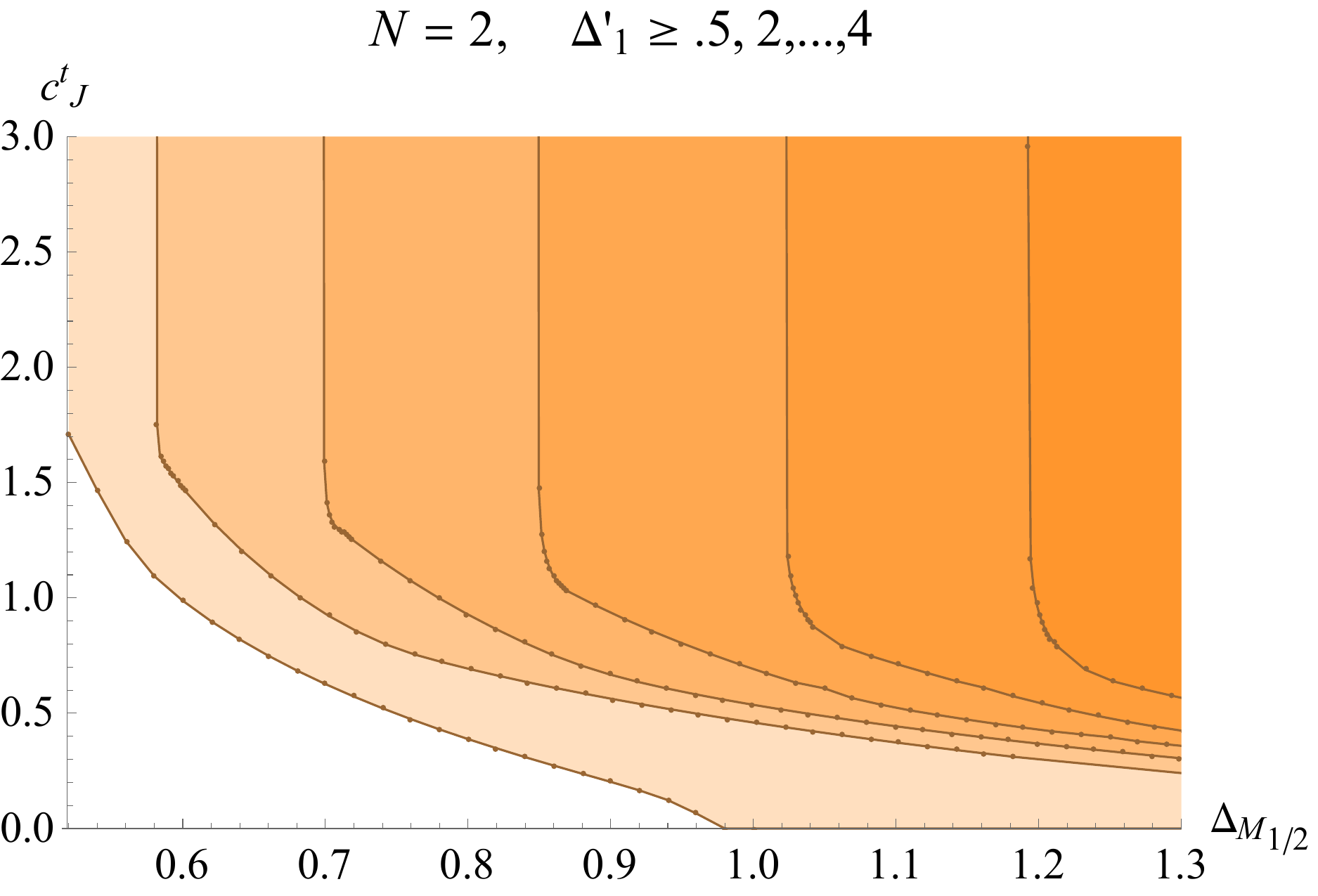}
 \includegraphics[width = 0.49\textwidth]{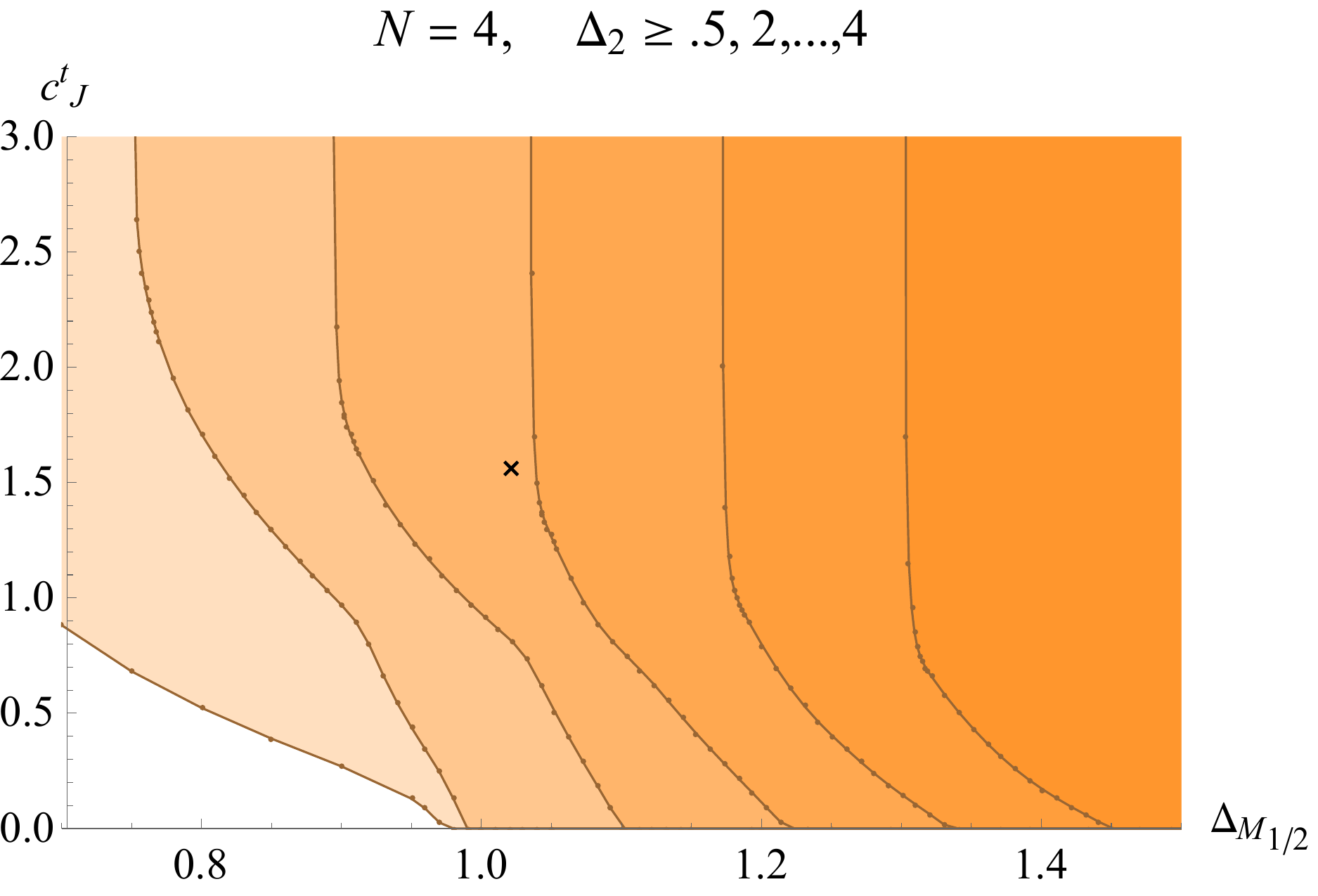} \\
 \includegraphics[width = 0.49\textwidth]{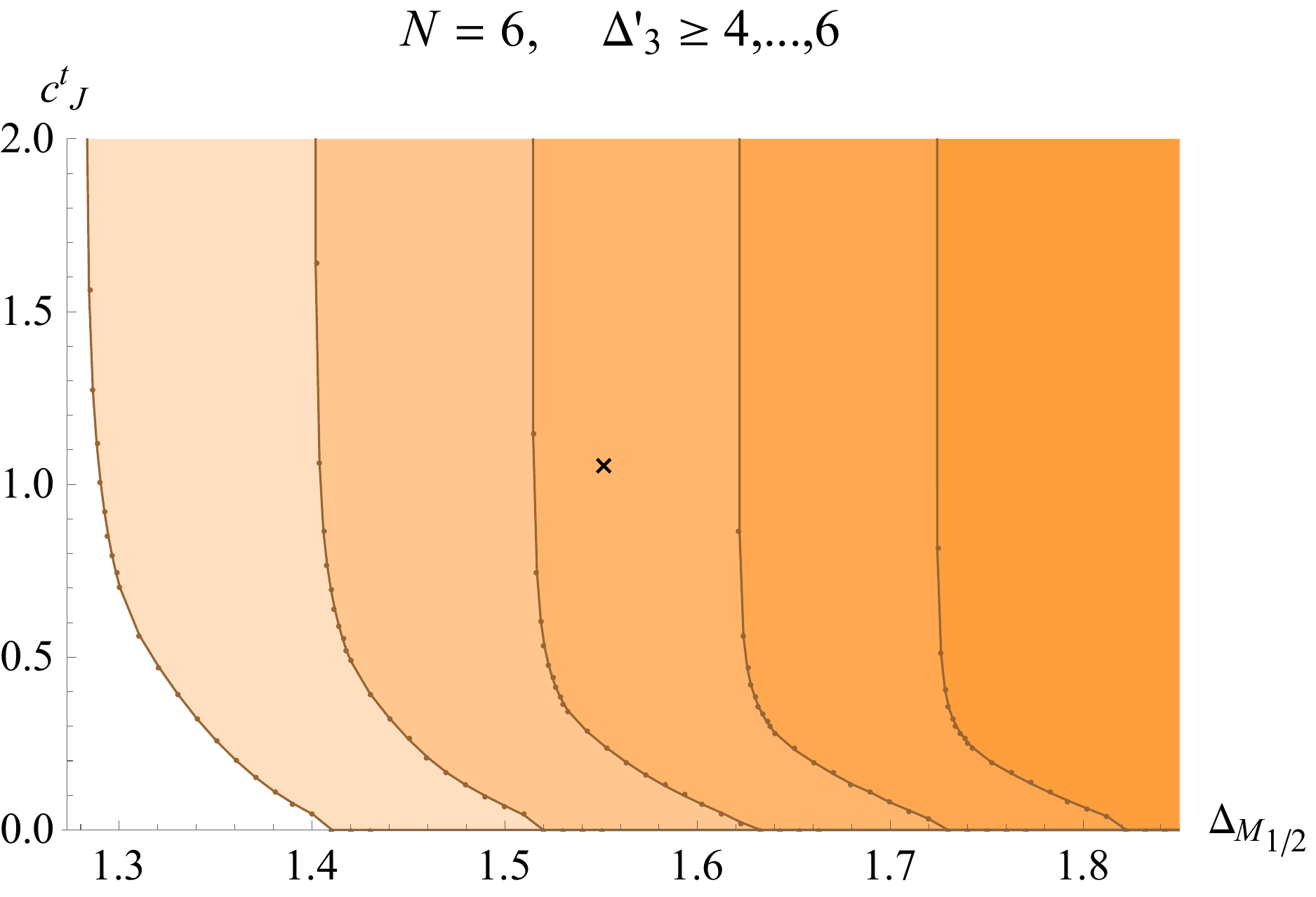}
  \caption{Bounds on topological $U(1)$ current charge $c_J^t$ in terms of basic $q=1/2$ monopole operator scaling dimension $\Delta_{M_{1/2}}$ in $d=3$ for $N=2,4,6$ with gaps $\Delta'_2\geq.5,2,2.5,3,3.5,4$ for $N=2$, $\Delta_4\geq.5,2,2.5,3,3.5,4$ for $N=4$, and $\Delta'_6\geq2,2.5,3,3.5,4$ for $N=6$ in the uncharged sector in the same $SU(N)$ representation $\left(2^{N/2}\right)$ as $M_1$. These bounds were computed with $\ell_\text{max}=25$ and $\Lambda=19$. The black crosses denote the large $N$ expansion values of $c_J^t$.  \label{fig:ctop}} \end{center}
 \end{figure}

\begin{figure}[ht!]
\begin{center}
 \includegraphics[width = 0.49\textwidth]{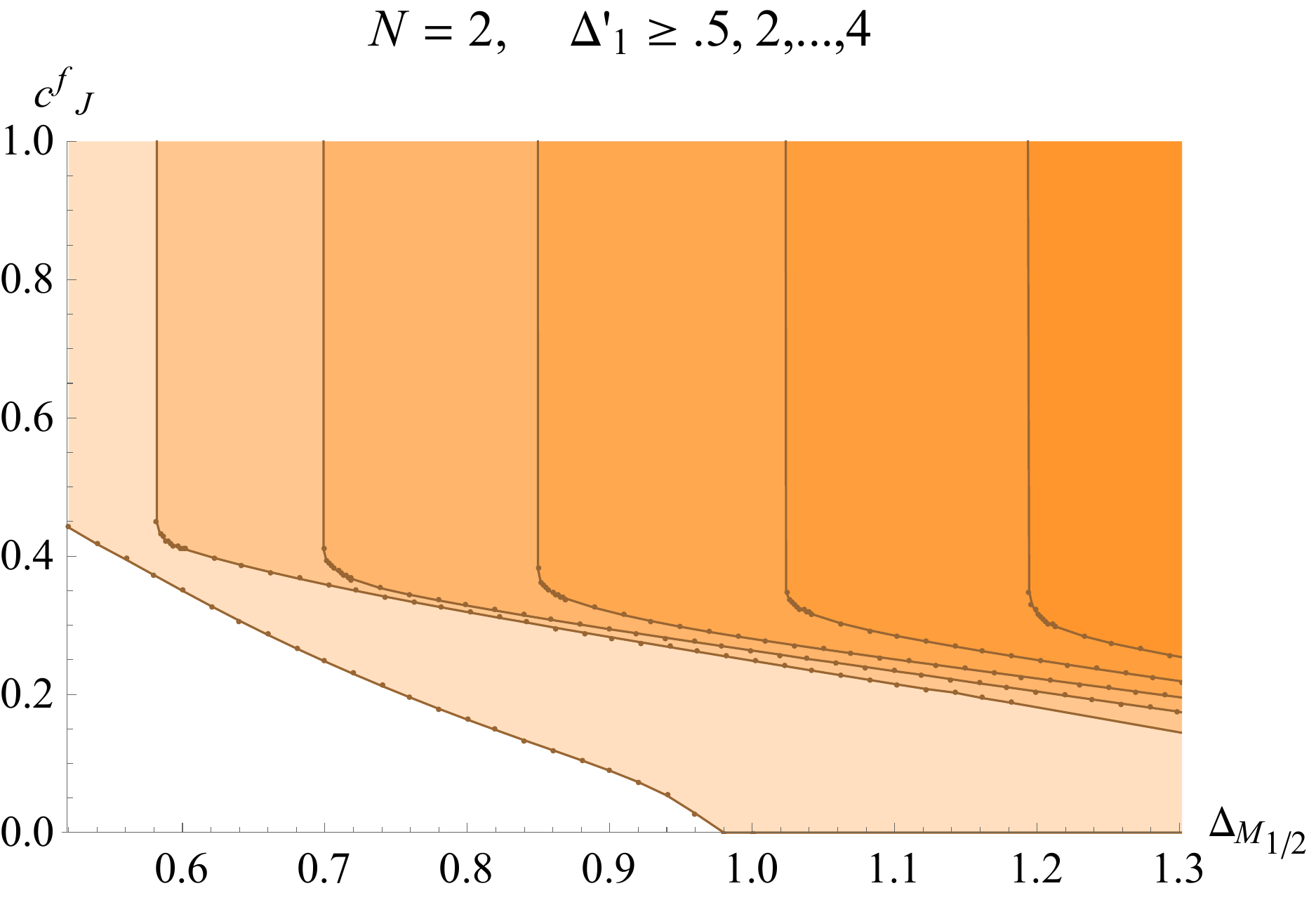}
 \includegraphics[width = 0.49\textwidth]{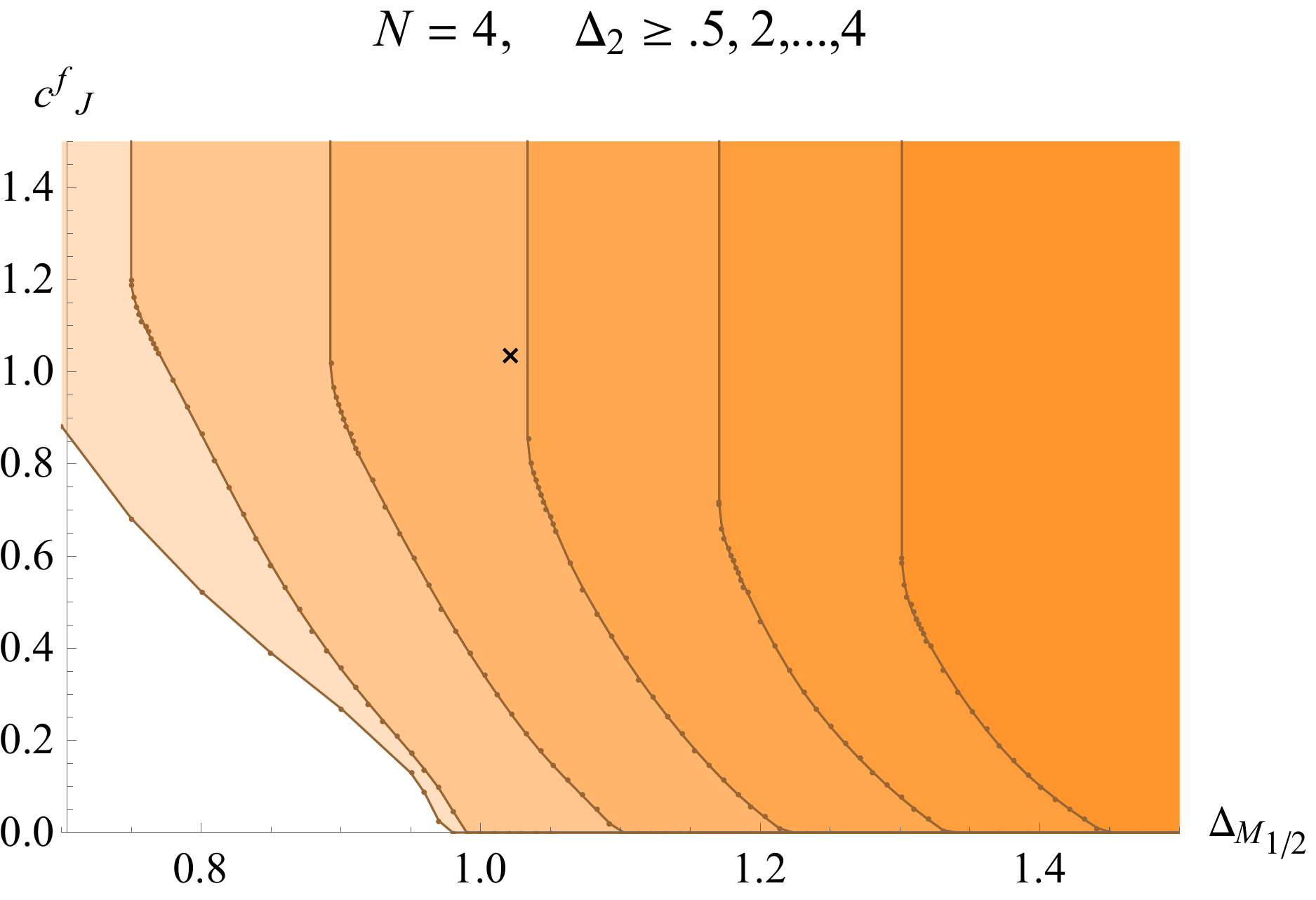} \\
 \includegraphics[width = 0.49\textwidth]{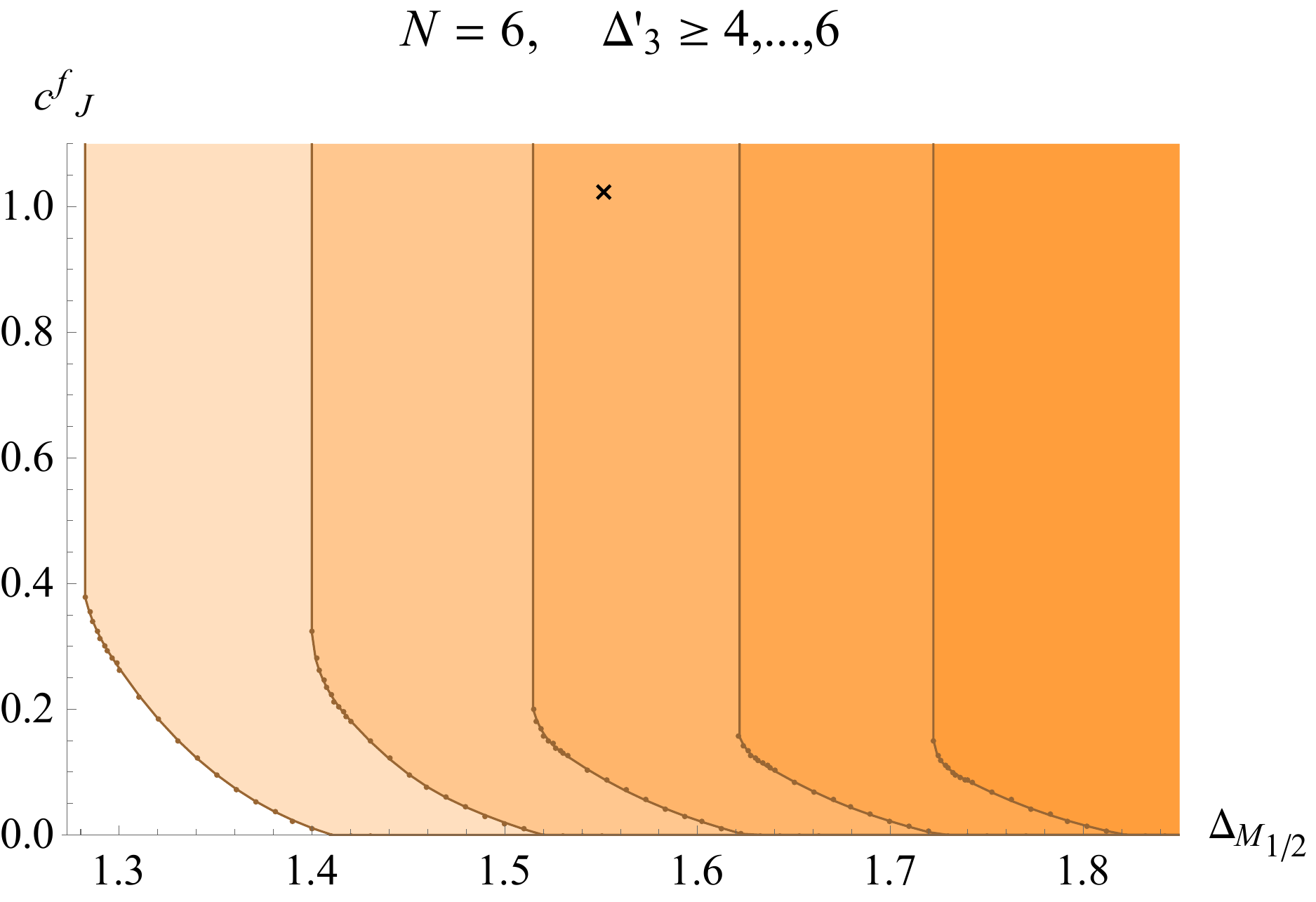}
 \caption{Bounds on $SU(N)$ flavor current charge $c_J^f$ in terms of basic $q=1/2$ monopole operator scaling dimension $\Delta_{M_{1/2}}$ in $d=3$ for $N=2,4,6$ with gaps $\Delta'_2\geq.5,2,2.5,3,3.5,4$ for $N=2$, $\Delta_4\geq.5,2,2.5,3,3.5,4$ for $N=4$, and $\Delta'_6\geq2,2.5,3,3.5,4$ for $N=6$ in the uncharged sector in the same $SU(N)$ representation $\left(2^{N/2}\right)$ as $M_1$. These bounds were computed with $\ell_\text{max}=25$ and $\Lambda=19$. The black crosses denote the large $N$ expansion values of $c_J^f$.  \label{fig:cf}} \end{center}
 \end{figure}

In Figures \ref{fig:cT}, \ref{fig:ctop}, and \ref{fig:cf}, we show bounds on the stress tensor central charge $c_T$, topological $U(1)$ current charge $c_J^t$, and $SU(N)$ flavor current charge $c_J^f$, respectively, plotted versus the basic monopole scaling dimensions $\Delta_{M_{1/2}}$. As with the bounds on $M_1$, we show these bounds for various gaps in the uncharged sector in the same $SU(N)$ representation $\left(2^{N/2}\right)$ as $M_1$. For $N=4,6$ we show the large $N$ values for $c_T$, $c_J^f$, and $c_J^t$ \eqref{currents}. The numerical bounds for $c_J^f$ and $c_T/N$ are not very restrictive, as they lie below the free theory value of $1$. On the other hand, the bound for $c_J^t$ is close to being saturated by the large $N$ expansion values for $N=4$. 

\section{Discussion}
\label{disc}
In this work, we studied constraints coming from crossing symmetry and unitarity in 3d CFTs with $SU(N) \times U(1)$ flavor symmetry that contain operators transforming as rank-$N/2$ anti-symmetric tensors of $SU(N)$ that have unit $U(1)$ charge.  An example of such a CFT is 3d QED, in which the most basic monopole operators transform under $SU(N) \times U(1)$ as above.  Interpreted in the context of 3d QED, we obtained bounds on the scaling dimension of the doubly-charged monopole operators in terms of the scaling dimension of the singly-charged one (Figures~\ref{fig:26} and~\ref{fig:4}), and also on the coefficients $c_T$, $c_J^t$, and $c_J^f$ appearing in the two-point function of the canonically normalized stress tensor, $U(1)$ flavor current, and $SU(N)$ flavor current (Figures~\ref{fig:cT},~\ref{fig:ctop}, and~\ref{fig:cf}).

We hope that our work represents the first steps toward a more systematic study of QED$_3$ using the conformal bootstrap.  We observed that when we impose certain gaps in the operator spectrum, we obtain a kink in our scaling dimension bounds (Figure~\ref{fig:4}) that is at the edge of an allowed region whose shape is similar to that seen in the study of theories with $\Z_2$ global symmetry.  In a further mixed correlator study, such a region turned into an island centered around the 3d Ising CFT, so it would be interesting to see if a mixed correlator study in the present setup would also lead to an island-shaped allowed region. In this study we also assumed that a CFT exists for all $N$, which is still an unsettled question. Perhaps by looking at mixed correlators one could exclude the existence of such a CFT for low $N$. We hope to report on such a mixed correlator study in an upcoming work.

\section*{Acknowledgments}

We thank Luca Iliesiu and David Simmons-Duffin for useful discussions.  This work was supported in part by the US NSF under Grant No.~PHY-1418069.

\clearpage

\appendix

\section{Crossing functions for $N=8,10,12,14$}
\label{moreCross}

The crossing functions given in \eqref{eq:crossingWithON} for the four-point function of Lorentz scalar operators in the $SU(N)$ irrep $\left(1^{N/2}\right)$ and the fundamental $SO(2)$ irrep require the input of the $SU(N)$ crossing functions, which differ with $N$. Below we list these functions for $N=8,10,12,14$, along with the signs $s_{R,n}$ defined in \eqref{VDefs}. When $N/2$ is even $s_{S,n}=s_{T,n}=s_{A,n}=s_{n}$, when $N/2$ is odd $s_{S,n}=-s_{T,n}=s_{A,n}=s_{n}$. The allowed spins are even in the following cases, and odd otherwise: $SO(2)$ irrep is $S,T$ and $SU(N)$ irrep $n$ is even, $SO(2)$ irrep is $A$ and $n$ odd. 
\es{SU8cross}{
&\raisebox{1cm}{$N=8:$} \vec{d}^{\,\mp,0}_{\Delta, \ell}= \begin{pmatrix}
7F_{\Delta,\ell}^\mp \\
0   \\
0  \\
 F_{\Delta,\ell}^\pm\\
0  \\
  \end{pmatrix}\,,
\qquad
 \vec{d}^{\,\mp,1}_{\Delta, \ell}=\begin{pmatrix}
0 \\
F_{\Delta,\ell}^\mp   \\
0  \\
 0\\
 F_{\Delta,\ell}^\pm  \\
  \end{pmatrix}\,,
\qquad
  \vec{d}^{\,\mp,2}_{\Delta, \ell}=\begin{pmatrix}
0 \\
0   \\
7F_{\Delta,\ell}^\mp  \\
-18 F_{\Delta,\ell}^\pm\\
21 F_{\Delta,\ell}^\pm\\
  \end{pmatrix}\,,  \\
%
& \vec{d}^{\,\mp,3}_{\Delta, \ell}=\begin{pmatrix}
 120F_{\Delta,\ell}^\mp \\
-4F_{\Delta,\ell}^\mp \\
-5F_{\Delta,\ell}^\mp \\
  -10 F_{\Delta,\ell}^\pm\\
  F_{\Delta,\ell}^\pm\\
  \end{pmatrix}\,,
  \qquad
    \vec{d}^{\,\mp,4}_{\Delta, \ell}=\begin{pmatrix}
-54F_{\Delta,\ell}^\mp \\
-15F_{\Delta,\ell}^\mp \\
-10F_{\Delta,\ell}^\mp \\
18 F_{\Delta,\ell}^\pm\\
15 F_{\Delta,\ell}^\pm\\
  \end{pmatrix}\,, \qquad
 s_n = \begin{pmatrix}
  1 & -1 & 1 & 1 & -1
 \end{pmatrix} \,.
}
\es{SU10cross}{
&\raisebox{1cm}{$N=10:$} \vec{d}^{\,\mp,0}_{\Delta, \ell}=\begin{pmatrix}
9F_{\Delta,\ell}^\mp \\
0   \\
0  \\
9 F_{\Delta,\ell}^\pm\\
0  \\
 0\\
  \end{pmatrix}\,,
\qquad
\vec{d}^{\,\mp,1}_{\Delta, \ell}=\begin{pmatrix}
0 \\
F_{\Delta,\ell}^\mp   \\
0  \\
0\\
 F_{\Delta,\ell}^\pm  \\
 0\\
  \end{pmatrix}\,,
 \qquad 
\vec{d}^{\,\mp,2}_{\Delta, \ell}=\begin{pmatrix}
0 \\
0   \\
9F_{\Delta,\ell}^\mp  \\
0\\
0  \\
 9 F_{\Delta,\ell}^\pm\\
  \end{pmatrix}\,, \\
& \vec{d}^{\,\mp,3}_{\Delta, \ell}=\begin{pmatrix}
 175F_{\Delta,\ell}^\mp \\
 -5 F_{\Delta,\ell}^\mp\\
-7 F_{\Delta,\ell}^\mp\\
-385 F_{\Delta,\ell}^\pm\\
23 F_{\Delta,\ell}^\pm\\
-35 F_{\Delta,\ell}^\pm\\
  \end{pmatrix}\,,
 \qquad
   \vec{d}^{\,\mp,4}_{\Delta, \ell}=\begin{pmatrix}
-625F_{\Delta,\ell}^\mp \\
0 \\
7F_{\Delta,\ell}^\mp \\
 -275F_{\Delta,\ell}^\pm \\
70 F_{\Delta,\ell}^\pm\\
-133 F_{\Delta,\ell}^\pm\\
  \end{pmatrix}\,,
\qquad
\vec{d}^{\,\mp,5}_{\Delta, \ell}=\begin{pmatrix}
-145F_{\Delta,\ell}^\mp   \\
-4F_{\Delta,\ell}^\mp   \\
-5F_{\Delta,\ell}^\mp   \\
385 F_{\Delta,\ell}^\pm  \\
-14 F_{\Delta,\ell}^\pm  \\
35 F_{\Delta,\ell}^\pm  \\
  \end{pmatrix}\,, \\
&  s_n = \begin{pmatrix}
 -1 & 1 & -1 & -1 & 1 & 1
 \end{pmatrix} \,.
}
\es{SU12cross}{
&\raisebox{1cm}{$N=12:$} \vec{d}^{\,\mp,0}_{\Delta, \ell}=\begin{pmatrix}
11F_{\Delta,\ell}^\mp \\
0   \\
0  \\
 0\\
11 F_{\Delta,\ell}^\pm\\
0  \\
 0\\
  \end{pmatrix}\,,
\qquad
\vec{d}^{\,\mp,1}_{\Delta, \ell}=\begin{pmatrix}
0 \\
3F_{\Delta,\ell}^\mp   \\
0  \\
 0\\
0\\
 F_{\Delta,\ell}^\pm  \\
 0\\
  \end{pmatrix}\,,
 \qquad
 \vec{d}^{\,\mp,2}_{\Delta, \ell}=\begin{pmatrix}
0 \\
0   \\
11F_{\Delta,\ell}^\mp  \\
 0\\
0\\
0  \\
 11 F_{\Delta,\ell}^\pm\\
  \end{pmatrix}\,, \\
&\vec{d}^{\,\mp,3}_{\Delta, \ell}=\begin{pmatrix}
0 \\
0  \\
0  \\
 3F_{\Delta,\ell}^\mp \\
 -180 F_{\Delta,\ell}^\pm\\
 20 F_{\Delta,\ell}^\pm\\
 -9 F_{\Delta,\ell}^\pm\\
  \end{pmatrix}\,, 
 \qquad
  \vec{d}^{\,\mp,4}_{\Delta, \ell}=\begin{pmatrix}
455F_{\Delta,\ell}^\mp \\
-70F_{\Delta,\ell}^\mp \\
-2F_{\Delta,\ell}^\mp \\
 -7F_{\Delta,\ell}^\mp \\
-175 F_{\Delta,\ell}^\pm\\
0\\
5F_{\Delta,\ell}^\pm\\
  \end{pmatrix}\,,
\qquad
\vec{d}^{\,\mp,5}_{\Delta, \ell}=\begin{pmatrix}
0   \\
-165F_{\Delta,\ell}^\mp   \\
0  \\
-21F_{\Delta,\ell}^\mp   \\
630 F_{\Delta,\ell}^\pm  \\
15 F_{\Delta,\ell}^\pm\\
-7 F_{\Delta,\ell}^\pm  \\
  \end{pmatrix}\,,\\
 & \vec{d}^{\,\mp,6}_{\Delta, \ell}=\begin{pmatrix}
2340F_{\Delta,\ell}^\mp   \\
-140F_{\Delta,\ell}^\mp   \\
7F_{\Delta,\ell}^\mp   \\
-35F_{\Delta,\ell}^\mp   \\
-1440 F_{\Delta,\ell}^\pm  \\
0 \\
-28 F_{\Delta,\ell}^\pm  \\
  \end{pmatrix}\,, 
 \qquad
 s_n = \begin{pmatrix}
 1 & -1 & 1 & 1 & 1& -1 & 1
\end{pmatrix} \,.
}
\es{SU14cross}{
&\raisebox{1cm}{$N=14:$}  \vec{d}^{\,\mp,0}_{\Delta, \ell}=\begin{pmatrix}
26F_{\Delta,\ell}^\mp \\
0   \\
0  \\
 0\\
26 F_{\Delta,\ell}^\pm\\
0  \\
 0\\
 0\\
  \end{pmatrix}\,,
\qquad
 \vec{d}^{\,\mp,1}_{\Delta, \ell}=\begin{pmatrix}
0 \\
22F_{\Delta,\ell}^\mp   \\
0  \\
 0\\
0\\
22 F_{\Delta,\ell}^\pm  \\
 0\\
 0\\
  \end{pmatrix}\,,
 \qquad
 \vec{d}^{\,\mp,2}_{\Delta, \ell}=\begin{pmatrix}
0 \\
0   \\
26F_{\Delta,\ell}^\mp  \\
 0\\
0\\
0  \\
 26 F_{\Delta,\ell}^\pm\\
 0\\
  \end{pmatrix}\,, \\
&\vec{d}^{\,\mp,3}_{\Delta, \ell}=\begin{pmatrix}
0 \\
0  \\
0  \\
 22F_{\Delta,\ell}^\mp \\
0\\
0  \\
 0\\
 22 F_{\Delta,\ell}^\pm\\
  \end{pmatrix}\,,
\qquad
 \vec{d}^{\,\mp,4}_{\Delta, \ell}=\begin{pmatrix}
11025F_{\Delta,\ell}^\mp \\
3465F_{\Delta,\ell}^\mp \\
55F_{\Delta,\ell}^\mp \\
 -99F_{\Delta,\ell}^\mp \\
-23625 F_{\Delta,\ell}^\pm\\
-11385 F_{\Delta,\ell}^\pm\\
-935 F_{\Delta,\ell}^\pm\\
-297 F_{\Delta,\ell}^\pm\\
  \end{pmatrix}\,,
\qquad
 \vec{d}^{\,\mp,5}_{\Delta, \ell}=\begin{pmatrix}
1029F_{\Delta,\ell}^\mp   \\
49F_{\Delta,\ell}^\mp   \\
-7F_{\Delta,\ell}^\mp   \\
3F_{\Delta,\ell}^\mp   \\
1323 F_{\Delta,\ell}^\pm  \\
1351 F_{\Delta,\ell}^\pm  \\
147 F_{\Delta,\ell}^\pm  \\
45 F_{\Delta,\ell}^\pm  \\
  \end{pmatrix}\,,\\
    & \vec{d}^{\,\mp,6}_{\Delta, \ell}=\begin{pmatrix}
8575F_{\Delta,\ell}^\mp \\
-735F_{\Delta,\ell}^\mp \\
-15F_{\Delta,\ell}^\mp \\
21_{\Delta,\ell}^- \\
-28175 F_{\Delta,\ell}^\pm\\
-10605 F_{\Delta,\ell}^\pm\\
-1065 F_{\Delta,\ell}^\pm\\
-357 F_{\Delta,\ell}^\pm\\
  \end{pmatrix}\,,
\qquad
 \vec{d}^{\,\mp,7}_{\Delta, \ell}=\begin{pmatrix}
-5775F_{\Delta,\ell}^\mp \\
255F_{\Delta,\ell}^\mp \\
-35F_{\Delta,\ell}^\mp \\
21_{\Delta,\ell}^- \\
-4725 F_{\Delta,\ell}^\pm\\
-6255 F_{\Delta,\ell}^\pm\\
-525 F_{\Delta,\ell}^\pm\\
-189 F_{\Delta,\ell}^\pm\\
  \end{pmatrix}\,, \\
  &s_n = \begin{pmatrix}
   -1 & -1 & - 1&  -1 & -1 & - 1& -1 & 1 
  \end{pmatrix} \,.
}

\bibliographystyle{ssg}
\bibliography{MonopolePaper}

\end{document}